\documentclass[runningheads]{llncs}
\usepackage[T1]{fontenc}
\usepackage{graphicx}
\usepackage{booktabs}
\usepackage[misc]{ifsym}
\newcommand{\corr}{(\Letter)}

\usepackage{amsmath}
\usepackage{amssymb}
\usepackage{mathtools}

\usepackage{amsthm}
\usepackage{enumitem}
\usepackage{newfloat}
\usepackage{makecell}
\usepackage{caption}
\DeclareCaptionType{listing}[Listing][List of Listings]

\usepackage[utf8]{inputenc}
\usepackage{listings}
\usepackage{xcolor}
\usepackage{tcolorbox}
\tcbuselibrary{listings, skins, breakable}

\definecolor{codegreen}{rgb}{0,0.6,0}
\definecolor{codegray}{rgb}{0.5,0.5,0.5}
\definecolor{codepurple}{rgb}{0.58,0,0.82}
\definecolor{backcolour}{rgb}{0.95,0.95,0.92}

\lstdefinestyle{mystyle}{
    language=Python,
    backgroundcolor=\color{backcolour},   
    commentstyle=\color{codegreen},
    keywordstyle=\color{magenta},
    numberstyle=\tiny\color{codegray},
    stringstyle=\color{codepurple},
    basicstyle=\ttfamily\footnotesize,
    breakatwhitespace=false,         
    breaklines=true,                 
    captionpos=b,                    
    keepspaces=true,                 
    numbers=left,                    
    numbersep=5pt,                  
    showspaces=false,                
    showstringspaces=false,
    showtabs=false,                  
    tabsize=2
}
\begin{document}

\title{QLLM: Do We Really Need a Mixing Network for Credit Assignment \\in Multi-Agent Reinforcement Learning?}

\titlerunning{QLLM: Do We Really Need a Mixing Network?} 



\author{Yuanjun Li\inst{1}\thanks{These authors contributed equally to this work.} \and
Zhouyang Jiang\inst{1}\protect\footnotemark[1] \and
Bin Zhang\inst{2} \and
Mingchao Zhang\inst{3} \and \\
Junhao Zhao\inst{1} \and
Zhiwei Xu\inst{1} \corr
}



\authorrunning{Y. Li et al.}



\institute{Shandong University, Jinan, China
\and
The Key Laboratory of Cognition and Decision Intelligence for Complex Systems, Institute of Automation, Chinese Academy of Sciences, Beijing, China
\and
Qilu University of Technology, Jinan, China \\
\email{liyuanjun@mail.sdu.edu.cn, zhiwei\_xu@sdu.edu.cn}
}

\maketitle              

\begin{abstract}
Credit assignment remains a fundamental challenge in multi-agent reinforcement learning (MARL) and is commonly addressed through value decomposition under the centralized training with decentralized execution (CTDE) paradigm.
However, existing value decomposition methods typically rely on predefined mixing networks that require additional training, often leading to imprecise credit attribution and limited interpretability.
We propose \textbf{QLLM}, a novel framework that leverages large language models (LLMs) to construct training-free credit assignment functions (TFCAFs), where the TFCAFs are nonlinear with respect to the global state and offer enhanced interpretability while introducing no extra learnable parameters.
A coder-evaluator framework is employed to ensure the correctness and executability of the generated code.
Extensive experiments on standard MARL benchmarks demonstrate that QLLM consistently outperforms baselines while requiring fewer learnable parameters. Furthermore, it demonstrates generalization across a broad set of value decomposition algorithms. Code is available at \url{https://github.com/MaoMaoLYJ/pymarl-qllm}.

\end{abstract}

\section{Introduction}

Recent advances in cooperative multi-agent reinforcement learning (MARL)~\cite{oroojlooy2023review} have led to widespread applications in domains such as assembly lines~\cite{kaven2024multi}, logistics distribution~\cite{yan2022reinforcement}, autonomous driving~\cite{zhang2024multi}, and swarm robotics~\cite{orr2023multi}. 
In such scenarios, multiple agents learn to interact with the environment collaboratively to accomplish designated cooperative tasks. 
Since agents are typically trained using a shared team-level reward, accurately attributing individual contributions is critical for effective policy optimization~\cite{oliehoek2016concise}. 
This challenge is commonly referred to as the \emph{credit assignment} problem, where inaccurate attribution can lead to suboptimal coordination behaviors, such as the emergence of lazy agents~\cite{liu2023lazy}.

To address the credit assignment problem in multi-agent systems, a variety of approaches have been explored in the literature. 
Policy gradient approaches~\cite{agarwal2021theory,iqbal2019actor,foerster2018counterfactual} address credit assignment through a centralized critic, where joint value or advantage functions are used to evaluate agents’ actions under shared rewards, and individual policies are updated based on agent-specific gradients derived from these centralized evaluations.
In addtion, value decomposition methods~\cite{sunehag2017value} have emerged as a dominant paradigm under the centralized training with decentralized execution (CTDE) framework~\cite{lowe2017multi}. 
These methods construct a joint value function that is decomposed into individual utility functions, ensuring that the optimal joint policy can be recovered from local components while enabling scalable and coordinated training. 
In practice, value decomposition methods typically rely on neural-network-based \emph{mixing networks}~\cite{rashid2020monotonic} to parameterize the credit assignment process. 
However, similar to other neural credit assignment approaches, they require additional training, incur substantial optimization overhead, and often lack semantic interpretability~\cite{linardatos2020explainable}.

With the rapid advancement of large language models (LLMs)~\cite{kasneci2023chatgpt}, they have demonstrated strong capabilities in natural language processing (NLP), including text understanding~\cite{bhattacharjee2023towards} and code generation~\cite{Jiang2024ASO}. 
Building on these capabilities, LLMs have increasingly been explored as general-purpose components in reinforcement learning (RL). 
In this context, LLMs can support RL in multiple ways, such as assisting reward design~\cite{ma2023eureka}, constructing abstract world models~\cite{poudel2023langwm}, or directly serving as decision-making policies~\cite{li2022pre}. 
By injecting structured prior knowledge and high-level reasoning into the learning process, LLMs complement conventional data-driven optimization in complex environments.

\begin{figure}[t]
\centering
\includegraphics[width= 0.75\linewidth]{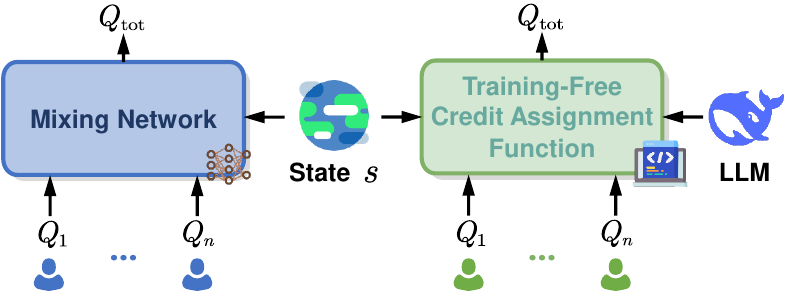}
\caption{Comparison between the traditional value decomposition method using the mixing network (\textbf{Left}) and our proposed novel paradigm QLLM leveraging LLMs (\textbf{Right}).}
\label{fig:example}
\end{figure}

Motivated by these developments, we investigate the use of LLMs to address the credit assignment problem in multi-agent systems.
This paper introduces \textbf{QLLM}, a novel value decomposition framework that eliminates the need for a learned mixing network, as shown in Figure~\ref{fig:example}.
To this end, a general \textit{coder-evaluator} framework is proposed to support reliable LLM-based code generation.
Within this framework, LLMs assume dual roles as both a code generator and a code evaluator~\cite{desmond2024evalullm}, enabling the synthesis and validation of executable functions.
Given task specifications and code generation instructions, the code generator produces a nonlinear \textbf{T}raining-\textbf{F}ree \textbf{C}redit \textbf{A}ssignment \textbf{F}unction (\textbf{TFCAF}), which serves as a direct substitute for the mixing network in value decomposition algorithms.
Because TFCAF is generated directly by LLMs, it requires no additional training and offers improved semantic interpretability, while remaining robust to complex state representations.
To further enhance reliability, a code verification mechanism is incorporated to ensure the correctness and executability of the generated code, thereby mitigating hallucination issues commonly observed in LLM-based code generation~\cite{martino2023knowledge}.
As a result, the proposed framework enables the construction of a dependable TFCAF without direct interaction with the environment, which can serve as a drop-in replacement for the mixing network and be readily integrated into a broad class of value decomposition methods.

The main contributions of this study are summarized as follows:
\begin{enumerate}
\item A coder-evaluator framework is adopted to enable the zero-shot construction of a reliable, training-free credit assignment function, termed \textbf{TFCAF}, via LLM-based code generation to the credit assignment problem in MARL.
\item A novel value decomposition framework, \textbf{QLLM}, is introduced by integrating TFCAF into MARL algorithms based on mixing network, allowing credit assignment without additional training or environment interaction.
\item Extensive experiments demonstrate that QLLM consistently outperforms existing credit assignment methods and provides improved semantic interpretability, particularly in complex state settings.
\end{enumerate}


\section{Related Work}
\subsection{Credit Assignment in MARL}
\label{subsec:CAMARL}

In recent years, the credit assignment problem has emerged as a fundamental challenge in multi-agent systems.
Policy-based approaches~\cite{yu2022surprising,zhang2024sequential} were among the first to introduce credit assignment mechanisms.
MADDPG~\cite{lowe2017multi} extends the DDPG~\cite{lillicrap2015continuous} framework to multi-agent settings by incorporating CTDE. 
This approach utilizes an actor-critic architecture~\cite{Konda1999ActorCriticA} with centralized critics, where the centralized critic has access to global information, enabling more effective learning in mixed cooperative-competitive scenarios. 
Similarly, COMA~\cite{foerster2018counterfactual} adopts a centralized critic and introduces a counterfactual baseline to estimate agent-specific advantage functions.
MAAC~\cite{iqbal2019actor} further enhances policy-based methods by introducing an attention mechanism into the critic, leading to more precise credit assignment.
However, most of these algorithms are not directly applicable to tasks with shared rewards.

Another widely adopted paradigm is the value decomposition approach~\cite{rashid2020monotonic,xu2022side}, which explicitly decomposes the globally shared return. 
As one of the earliest methods in this category, VDN~\cite{sunehag2017value} models the global value function as a linear summation of individual value functions. 
While simple and computationally efficient, VDN struggles to capture complex cooperative relationships due to its linear decomposition. 
QMIX~\cite{rashid2020monotonic} improves upon VDN by using a monotonic mixing network to non-linearly combine individual value functions, allowing for more expressive representations. 
However, the assumption of monotonicity between individual and global values may restrict its flexibility in highly complex environments. 
QTRAN~\cite{son2019qtran} and QPLEX~\cite{wang2020qplex} mitigate this limitation by relaxing the monotonicity constraint, enabling the model to learn richer representations, albeit at the cost of increased computational complexity.
RIIT~\cite{hu2021rethinking} goes further by modeling implicit influence relationships between agents, enabling better credit assignment in tasks with long-term dependencies. 
Nevertheless, RIIT ultimately retains a mixing network structure analogous to that used in QMIX, in order to establish the mapping from individual agent Q-values to the global Q-value.
While the mixing network plays a central role in implementing credit assignment, it often suffers from limited interpretability and slow convergence during training.
As a result, despite the success of value decomposition methods, achieving efficient, precise, and interpretable credit assignment in multi-agent systems remains an open challenge.

\subsection{LLM-Enhanced RL}
	
Leveraging their large-scale pretraining and strong generalization capabilities, LLMs have been increasingly integrated into RL to enhance performance in multitask learning, sample efficiency, and high-level task planning~\cite{zhang2023controlling,li2024verco}. 
Recent studies have explored the use of LLMs to generate structured decision policies or programmatic representations from task descriptions, enabling more interpretable and adaptable control strategies~\cite{deng2024new,li2022pre}.  
In addition, LLMs can provide natural language rationales for agent behaviors, improving the transparency and explainability of decision-making processes~\cite{das2023state2explanation,puiutta2020explainable}. 
LLMs have been investigated as tools for shaping reward signals, where automatically generated reward functions or feedback mechanisms help guide agents toward more effective behaviors~\cite{ma2023eureka,ma2025vision}. 
Collectively, these works highlight the potential of LLMs as flexible components for enhancing learning signals and structural priors in RL.
Recent work has begun to explore the use of LLMs for temporal and agent-level credit assignment in MARL.
Qu et al.~\cite{qu2025latent} proposed latent rewards to capture cooperative relationships, effectively addressing credit assignment under sparse rewards and outperforming existing baselines. Meanwhile, Lin et al.~\cite{lin2025speaking} employed LLMs to learn potential-based reward functions across multiple queries, allowing for more accurate estimation of individual agent contributions.
Despite their effectiveness, these methods primarily focus on reward decomposition and remain limited to settings compatible with single-agent reinforcement learning formulations, which restricts their scalability and applicability to general MARL algorithms.
To overcome these limitations, QLLM is introduced as a multi-agent credit assignment framework that uses LLM-based code generation to construct training-free credit assignment functions compatible with value decomposition methods.


\section{Preliminaries}
	
\subsection{Dec-POMDP}

In this work, we focus on a fully cooperative MARL setting formalized as a decentralized partially observable Markov decision process (\mbox{Dec-POMDP})~\cite{oliehoek2016concise}.
A Dec-POMDP is typically defined by a tuple \(G = \langle S, A, P, r, Z, O, n, \gamma \rangle\). 
In this setting, \(n\) agents indexed by \(i \in \{1, \dots, n\}\) operate in an environment with a true underlying state \(s \in S\). 
At each timestep, agent \(i\) selects an action \(a^i \in A\) based solely on its private observation \(z_i \in Z\), which is sampled according to the observation function \(O(s, i)\). 
The joint action of all agents is denoted by \(\boldsymbol{a} \in \boldsymbol{A}\).
The environment evolves according to the transition probability function \(P(s^\prime \mid s, \boldsymbol{a}): S \times \boldsymbol{A} \to [0,1]\), which specifies the probability of transitioning to state \(s^\prime\) given the current state \(s\) and joint action \(\boldsymbol{a}\). 
All agents cooperate to maximize a shared reward signal defined by the reward function \(r(s, \boldsymbol{a}): S \times \boldsymbol{A} \to \mathbb{R}\). 
The objective in this setting is to learn joint policies that maximize the expected cumulative discounted reward \(\sum_{j=0}^\infty \gamma^j r_{t+j}\), where \(\gamma \in [0, 1)\) is a discount factor.
	
Dec-POMDPs pose inherent challenges due to partial observability and non-stationarity.
Beyond these difficulties, a central challenge in Dec-POMDPs is the \emph{credit assignment} problem, which is the primary focus of this work.
Because agents typically optimize a shared team reward, accurately attributing individual contributions to the collective outcome is non-trivial and plays a critical role in effective coordination and learning.

\subsection{Value Decomposition and IGM Principle}

Traditional value decomposition methods employ a mixing network to aggregate the local action-value functions \(Q_i\) of individual agents into a global action-value function \(Q_{\text{tot}}\) following the Individual-Global-Maximum (IGM) principle~\cite{rashid2020monotonic},whereby the joint action that maximizes \( Q_{\text{tot}} \) must correspond to the combination of each agent’s individually optimal action.
Formally, the IGM principle can be expressed as:
\begin{equation}
\arg \max _{\boldsymbol{a}} Q_{\mathrm{tot}}(\boldsymbol{\tau}, \boldsymbol{a})=\left(\begin{array}{c}
\arg \max _{a^{1}} Q_{1}\left(\tau^{1}, a^{1}\right) \\
\vdots \\
\arg \max _{a^{n}} Q_{n}\left(\tau^{n}, a^{n}\right)
\end{array}\right),
\end{equation}
where \(\boldsymbol{\tau} = (\tau^1, \dots, \tau^n)\) represents the joint action-observation histories of all agents.
This equation implies that the maximization of the global Q-value function is aligned with the maximization of each agent’s local Q-value function, enabling decentralized execution while preserving global optimality. 
Building upon this principle, Qatten~\cite{yang2020qatten} provides a theoretical extension, showing that \(Q_\text{tot}\) can be expressed as a linear combination of the individual \(Q_i\) functions:
\begin{equation}
Q_\text{tot}(s, \boldsymbol{a}) = \sum_{i=1}^n \sum_{h=1}^H w_{i,h}(s) Q_i(\tau^i, a^i) + b. 
\label{eq:qatten}
\end{equation}
In this formulation, \(w_{i,h}(s)\) denotes a coefficient that depends on the state \(s\), and \(H\) is the number of attention heads. 
Theoretically, these coefficients can be derived from a higher-order functional expansion of \(Q_\text{tot}\) with respect to \(Q_i\), where the contribution of each term diminishes rapidly with increasing derivative order. 
This indicates that the influence of higher-order interaction terms on \(Q_\text{tot}\) is sufficiently small to be ignored or effectively absorbed into a bias term \(b\), allowing for a simplified model formulation and reduced computational complexity.


\section{Methodology}

\begin{figure*}[htbp]
\centering
\includegraphics[width=\textwidth]{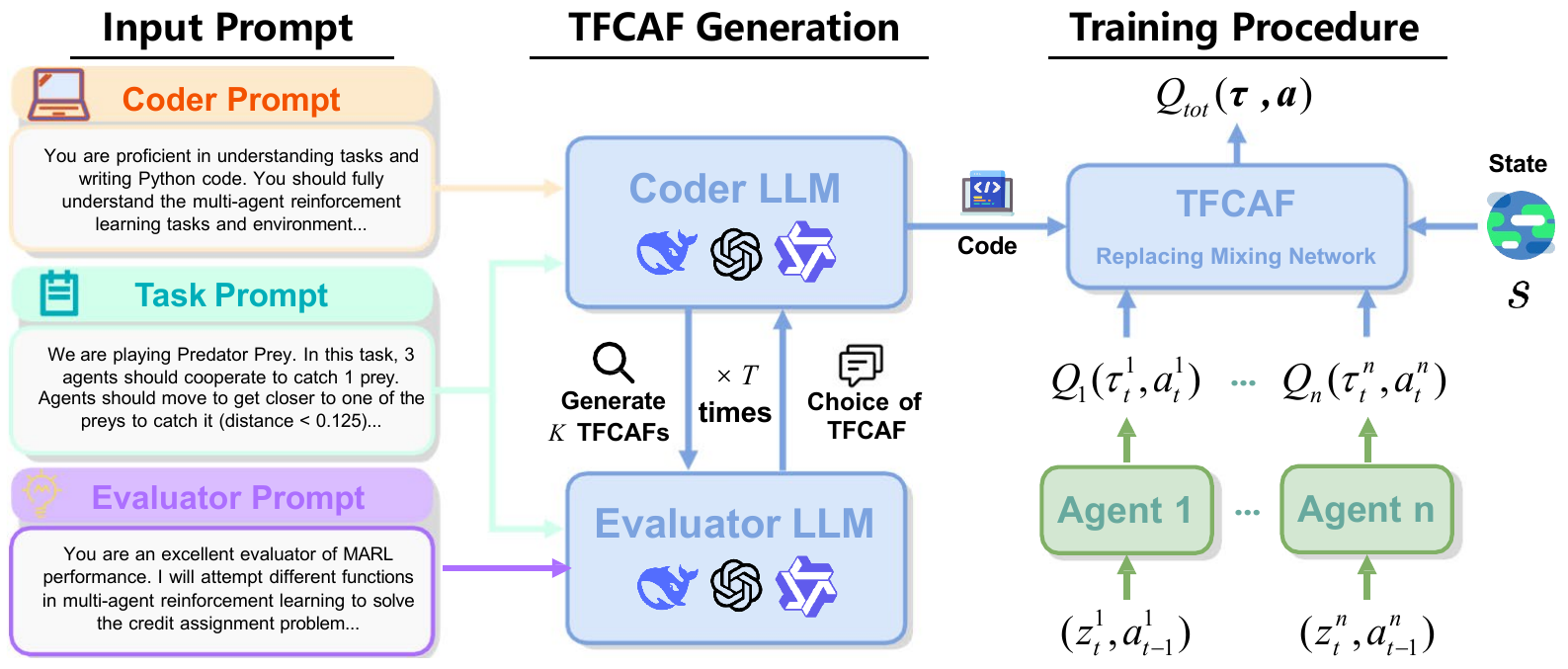}
\caption{QLLM framework: \( M_{\text{coder}} \) generates candidate TFCAFs, and \( M_{\text{evaluator}} \)  selects the optimal function after automated verification.}
\label{fig:framework}
\end{figure*}

This section presents a detailed discussion of the implementation of each component of QLLM, whose overall framework is illustrated in Figure~\ref{fig:framework}.
We first introduce the concept of TFCAF, a training-free credit assignment function capable of accurately assigning credit to individual agents. 
Leveraging the generative capabilities of LLMs, TFCAF is constructed through prompt-based interaction with LLMs.
However, when applied to complex tasks, LLMs typically encounter two fundamental challenges, hallucination~\cite{tonmoy2024comprehensive} and a lack of rigorous reasoning. 
These limitations motivate the design of our proposed \textit{coder-evaluator} framework. 
By integrating these components, QLLM is able to generate highly accurate TFCAF functions for multi-agent credit assignment, providing an effective alternative to the traditional mixing networks used in value decomposition approaches.

\subsection{Training-Free Credit Assignment Function}  

Traditional mixing networks often rely on hypernetwork~\cite{HyperNetworks} to generate state-dependent parameters. However, such neural-network-based credit assignment functions cannot effectively incorporate prior knowledge, which may result in slower convergence and reduced interpretability.

To address these issues, we propose a training-free credit assignment function (TFCAF), which serves as the core module of QLLM.
This module is generated by LLMs based on a series of task-related prompts, with correctness ensured through structured code verification.
As illustrated in Equation~\eqref{eq:qatten}, the global Q-value function \( Q_{\text{tot}} \) can theoretically be decomposed into a weighted sum of individual local Q-values with an additional bias term.
Motivated by this decomposition paradigm, a similar formulation is adopted in the design of TFCAF.
Its mathematical expression is as follows:
\begin{equation}
Q_{\text{tot}}\left( s, \boldsymbol{a} \right) =\sum_{i=1}^n{f_{w}^{i}}\left( s \right) Q_i\left( \tau^i, a^i \right) +f_{b}\left( s \right) .
        \label{eq:qllm}
\end{equation}

The coefficient \( f_{w}^{i}\left( s \right) \) is a state-dependent weight, and \( f_{b}\left( s \right) \) is a bias term that also depends on the global state.
Both are key components of the TFCAF, which are generated by LLMs. A detailed theoretical proof justifying this decomposition formulation is provided in Appendix~\ref{sec:appendix_proof}. 
In the following, we provide a detailed explanation of how the TFCAF is generated, as well as how it facilitates credit assignment in multi-agent systems.

\subsection{Coder-Evaluator Framework}

Our framework employs two distinct LLM roles: the \textit{coder LLM} \(M_{coder}\) for generating candidate TFCAFs and the \textit{evaluator LLM} \(M_{evaluator}\) for critical selection and validation, ensuring the robustness of the generated functions with minimal human involvement.

The interaction between the two models is driven by two types of prompts. 
Specifically, \textit{Task prompts} (\(P_{task}\)) provides environment-specific details, while \textit{Role prompts} (\(P_{role}\)) define the fixed responsibilities and operational roles of each LLM within the framework. 
These role prompts are further categorized into \textit{coder prompt} and \textit{evaluator prompt}, which remain unchanged across different RL tasks and correspond to the Coder LLM and the Evaluator LLM, respectively.
This separation between task-specific and role-specific prompts enables QLLM to adapt to new cooperative tasks with minimal prompt modification, without requiring extensive manual tuning or subjective rewriting.
Detailed prompt templates used in the framework are provided in Appendix~\ref{sec:A}.

The generation of TFCAF can be formally described through the following three clearly-defined steps.


\noindent\textbf{Initial Generation.} 
At the initial stage, the Coder LLM \(M_{\text{coder}}\) generates a set of candidate credit assignment functions conditioned on the provided prompts.
Specifically, given the role-specific prompts \(P_{\text{role}}\) and the task-specific prompts \(P_{\text{task}}\), \(M_{\text{coder}}\) produces \(K\) candidate functions:
\begin{equation}
\phi_1, \phi_2, \dots, \phi_K \ \Leftarrow \ M_{\text{coder}}\big(P_{\text{role}}, P_{\text{task}}\big),
\end{equation}
where each \(\phi_k\) represents a candidate TFCAF.
Each generated function explicitly specifies how to compute the state-dependent weight functions \(\{f_w^i(s)\}_{i=1}^n\) and the bias term \(f_b(s)\) in Equation~\eqref{eq:qllm}, typically in the form of executable code or programmatic expressions.
Consequently, each \(\phi_k\) defines a complete mapping from the global state \(s\) to the mixing coefficients used for credit assignment, without introducing additional learnable parameters.

\noindent\textbf{Error Detection and Correction.}
Due to the inherent tendency of LLMs to produce hallucinated or imperfect code, the generated candidate functions may contain syntax errors, dimensional mismatches, or invalid operations.
To ensure executability, each candidate TFCAF is first compiled and executed once using a set of available input states and corresponding local action-value estimates.
If an execution error occurs, the corresponding error message returned by the compiler or runtime environment is collected and provided as structured feedback to the Coder LLM \(M_{\text{coder}}\).
Conditioned on this feedback, \(M_{\text{coder}}\) regenerates a corrected version of the faulty TFCAF:
\begin{equation}
\phi_{\text{correct}} \ \Leftarrow \ M_{\text{coder}}\big(P_{\text{role}},\, P_{\text{task}},\, \text{error}\big).
\end{equation}
This procedure continues until all \(K\) candidate TFCAFs execute successfully, ensuring basic syntactic and execution-level correctness before they are passed to subsequent evaluation stages.

\noindent\textbf{Evaluation of TFCAF.} 
After execution-level validation, the Evaluator LLM \(M_{\text{evaluator}}\) evaluates the candidate TFCAFs to identify the most suitable one for credit assignment.
Given the validated candidate set \(\Phi = \{\phi_1, \phi_2, \dots, \phi_K\}\), \(M_{\text{evaluator}}\) focuses on whether a TFCAF provides a coherent and task-aligned mapping from the global state to agent-wise credit weights, rather than relying on empirical performance metrics.

The evaluation and final choice are fully guided by structured role and task prompts and are performed automatically by \(M_{\text{evaluator}}\), without manual ranking or subjective human intervention:
\begin{equation}
\phi_{\text{choice}} \ \Leftarrow \ M_{\text{evaluator}}\big(P_{\text{role}},\, P_{\text{task}},\, \Phi \big),
\end{equation}
where the selected TFCAF is subsequently adopted as the credit assignment function in QLLM. Subsequently, $\phi_{\text{choice}}$ is fed back into $M_{\text{coder}}$ to serve as a reference for the next cycle of generation, and QLLM iteratively repeats this process for a total of $T$ rounds to ensure continuous logic refinement.

\subsection{Training Procedure}
With the \textit{coder-evaluator} framework, we obtain a training-free credit assignment function. 
Like QMIX and VDN, QLLM is a value-based MARL method developed under the CTDE paradigm.
Its objective is to learn a global action-value function \(Q_{\text{tot}}(s, \boldsymbol{a})\) that can be decomposed into individual utility functions \(Q_i(\tau^i, a^i)\), while avoiding the constraints imposed by conventional hypernetwork-based mixing networks. 
This design enables the LLM to model more complex and nonlinear interactions between local and global Q-values, allowing it to better adapt to the specific structural properties of the task.

As with most value decomposition approaches, each agent \(i\) in QLLM employs a deep neural network to approximate its individual action-value function \(Q_i(\tau^i, a^i; \theta)\), where \(\theta\) denotes the set of learnable parameters. 
These individual Q-values are then aggregated by the TFCAF, which is automatically generated by LLMs, to produce the global action-value function:
\begin{equation}
Q_{\text{tot}}(s, \boldsymbol{a}) = f_{\text{TFCAF}}(Q_1, \dots, Q_n; s),
\end{equation}
where \(f_{\text{TFCAF}}\) is a parameterized function in which the weights are allowed to take arbitrary values, while the biases are also unconstrained. 
\(f_{\text{TFCAF}}\) receives the individual action-value functions \(Q_i\) as input and utilizes the global state \(s\) to generate the corresponding weights \(f_{w}^{i}\) and bias \(f_{b}\) through the coder-evaluator framework, respectively, both of which are modeled as nonlinear functions of the global state.
Notably, all parameters in the TFCAF are generated by LLMs and remain fixed during training. The global Q-value is then computed as shown in Equation~\eqref{eq:qllm}.
The training objective is to minimize the temporal-difference (TD) loss between the predicted total Q-value and the target Q-value:
\begin{equation}
\mathcal{L}(\theta) = \mathbb{E}_{(s, \boldsymbol{a}, r, s')} \bigg[ \big( y - Q_{\text{tot}}(s, \boldsymbol{a};\theta) \big)^2 \bigg],
\end{equation}
where the target value \(y\) follows the Bellman equation:
\begin{equation}
y = r + \gamma \max_{\boldsymbol{a}'} Q_{\text{tot}}(s', \boldsymbol{a}';\theta^-),
\end{equation}
where $\theta^-$ denotes the parameters of the target agent network.

Compared to other neural network-based approaches for multi-agent credit assignment, QLLM involves fewer trainable parameters. 
This facilitates more precise credit allocation among agents in the early training stages, thereby accelerating the convergence of their decision-making policies. 
Meanwhile, the incorporation of TFCAF allows QLLM to effectively exploit prior knowledge while employing an interpretable credit assignment mechanism.


\section{Experiments}

To comprehensively evaluate the efficacy and versatility of the proposed QLLM framework, we design and conduct experiments focused on the following objectives:
(1) assessing the performance of QLLM across a diverse range of cooperative multi-agent environments; 
(2) validating the generality and compatibility of QLLM by integrating it with various established value decomposition algorithms; 
(3) evaluating its robustness to high-dimensional global state spaces by systematically increasing the number of agents and state dimensionality; and
(4) investigating the interpretability of TFCAF to provide insights into the generated credit assignment logic.
Furthermore, additional ablation studies and detailed comparisons of the number of learnable parameters across different algorithms are provided in Appendix~\ref{sec:B}.

\subsection{Experimental Setups}
We tested QLLM's performance on four common MARL benchmarks and compared it with several well-known baseline credit assignment algorithms. The primary LLM used in our method is the \textbf{DeepSeek-R1} \cite{guo2025deepseek} inference model. For most experiments, the reported values represent the mean over five independent training runs, with 90\% confidence intervals indicated by shaded areas. Notably, for SMAC scenarios, the solid lines represent the median test win rate, while the shaded regions denote the 25th-75th percentile interquartile range. Our experimental environments include:  
(1) Level-Based Foraging (LBF)~\cite{papoudakis2020benchmarking}, for testing in sparse reward settings; 
(2) Google Research Football (GRF)~\cite{kurach2020google}, to assess performance in complex, realistic environments;
(3) Multi-agent Particle Environments (MPE)~\cite{lowe2017multi}, for evaluating performance in high-dimensional state spaces; and 
(4) StarCraft Multi-Agent Challenge (SMAC)~\cite{samvelyan2019starcraft}, to test performance in complex and highly cooperative multi-agent scenarios.
    
We compare QLLM with seven baseline algorithms:  
QMIX~\cite{rashid2020monotonic}, QPLEX~\cite{wang2020qplex}, Qatten~\cite{yang2020qatten}, RIIT~\cite{hu2021rethinking}, COMA~\cite{foerster2018counterfactual}, Weighted QMIX~\cite{rashid2020weighted} and MASER~\cite{jeon2022maser}.   
A detailed introduction and implementation details for all baselines and environments are provided in Appendix~\ref{sec:C}.

\subsection{The Superiority of QLLM}
\begin{figure*}[htbp]
    \centering
    \includegraphics[width=\textwidth]{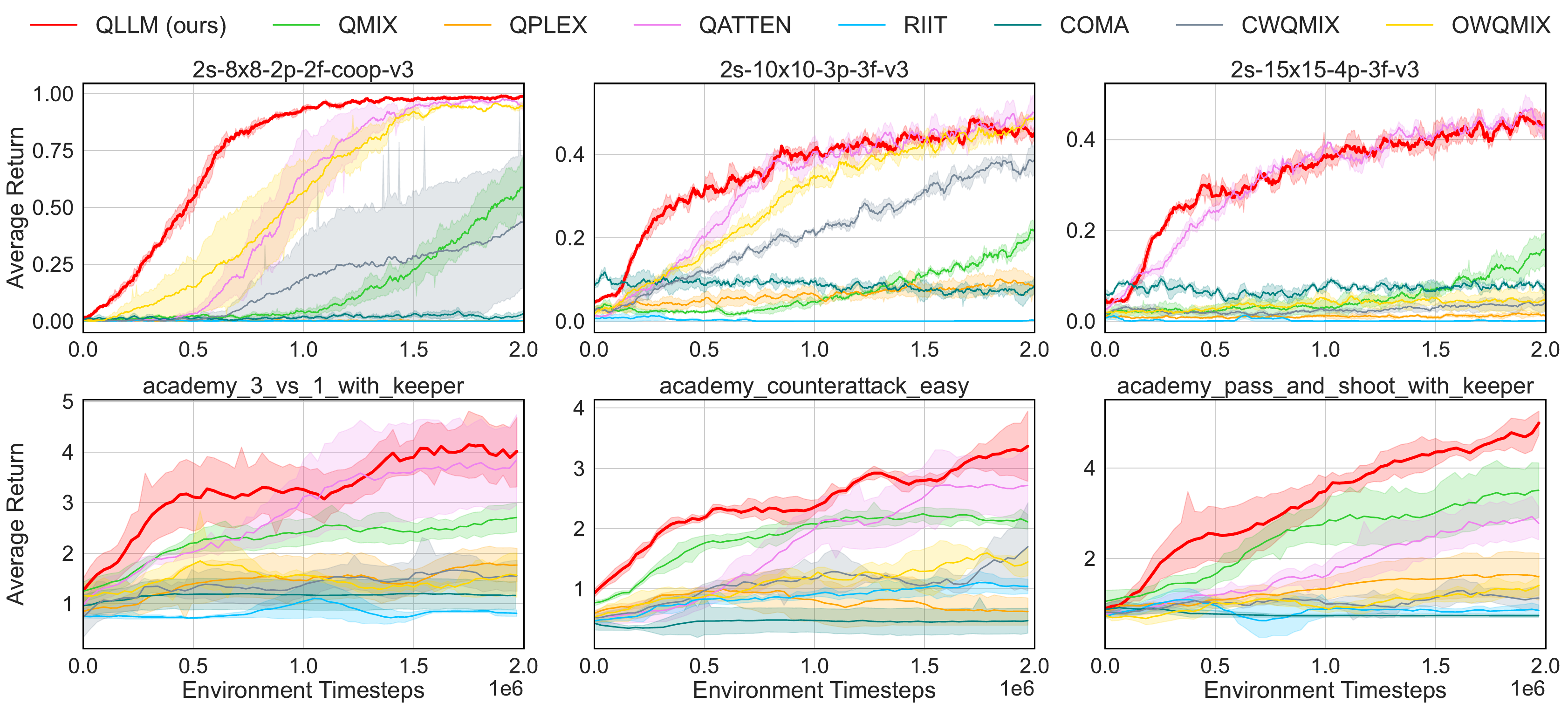}
    \caption{Performance comparison in LBF and GRF environments.
    }
    \label{fig:main_ex}
\end{figure*}

Figure~\ref{fig:main_ex} show the performance of various algorithms in LBF and GRF environments. It is evident that the QLLM algorithm outperforms the baseline in both the LBF environment and the GRF environment.

\subsection{The Generality of QLLM}
\begin{figure*}[t]
    \centering
    \includegraphics[width=\textwidth]{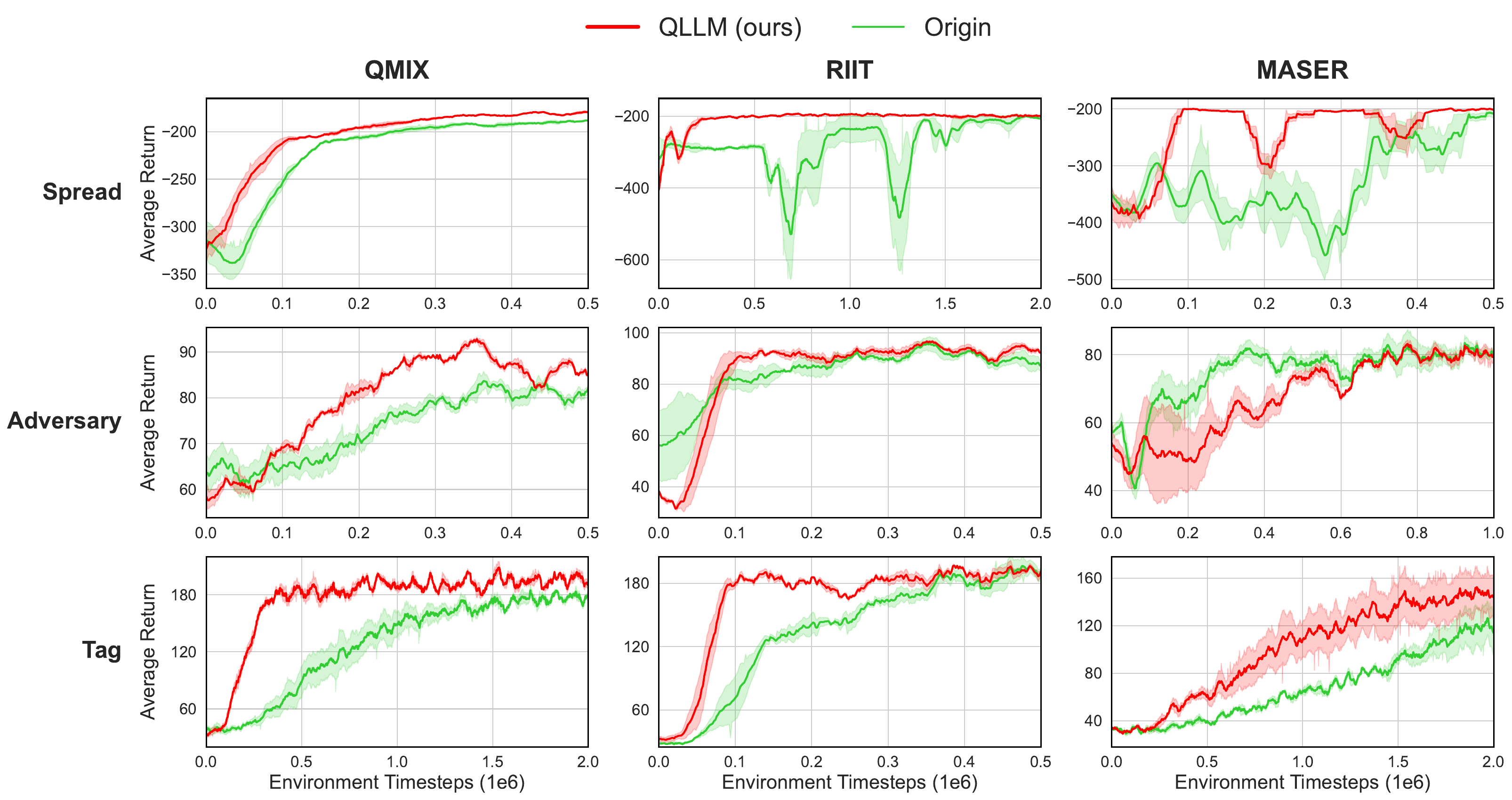}
    \caption{Comparison of QLLM-enhanced variants with their original baselines in the MPE environments.
    }
    \label{fig:test}
\end{figure*}

QLLM is based on value decomposition methods, so it is compatible with most MARL algorithms that use mixing networks to compute global Q-value. To verify this compatibility, we extend our evaluation beyond QMIX, using the TFCAF generated by QLLM to replace the mixing networks in two other value decomposition algorithms: RIIT and MASER. We conducted experiments in the MPE environment, and the experimental results are shown in Figure~\ref{fig:test}. Our results demonstrate that QLLM can be integrated with various value decomposition algorithms to significantly enhance their performance.

\subsection{The Performance of QLLM in High Dimensional State Spaces}
\begin{figure}[t]
    \centering
    \includegraphics[width=\textwidth]{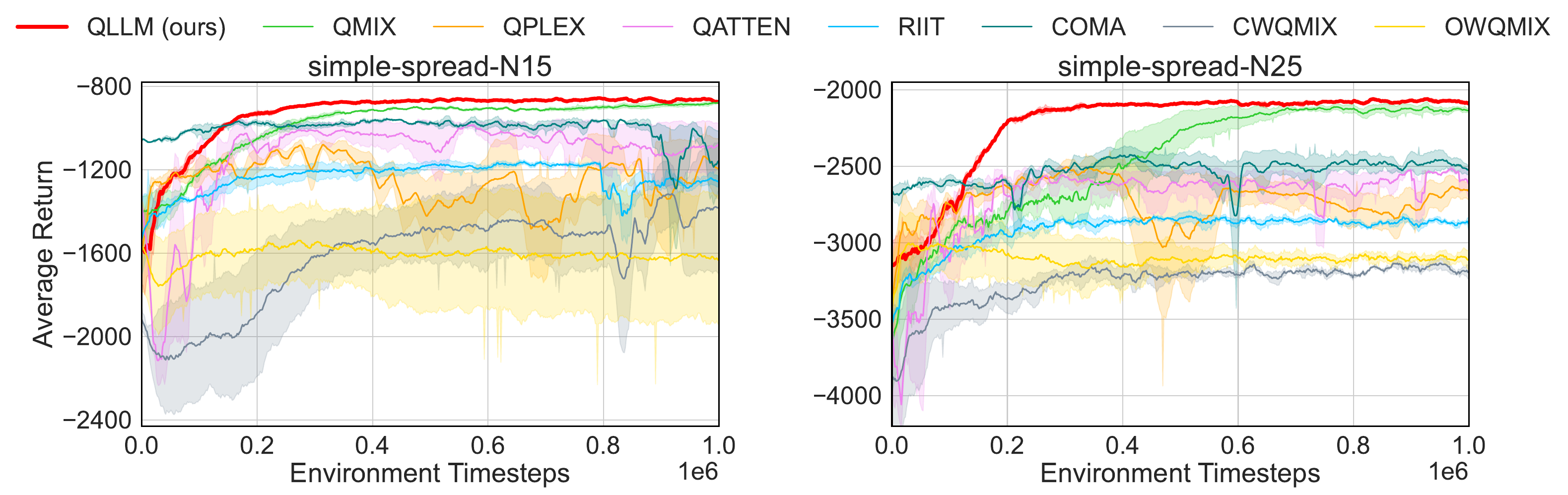}
    \caption{The results of high-dimensional testing in \textit{simple-spread} task.}
    \label{fig:high}
\end{figure}

We evaluated the scalability of QLLM by increasing the number of agents to $N=15$ and $N=25$ in the \textit{simple-spread} task. As shown in Figure~\ref{fig:high}, as state dimensionality grows, credit assignment becomes difficult for traditional value decomposition baselines, leading to significant performance degradation. In contrast, QLLM maintains high contribution assignment accuracy. This robustness stems from the LLM-generated nonlinear TFCAFs, which leverage well-defined task logic that is invariant to dimensionality, unlike traditional mixing network that struggle with high-dimensional parameter optimization.

\subsection{The Performance on SMAC}
\begin{figure}[t]
    \centering
    \includegraphics[width=\textwidth]{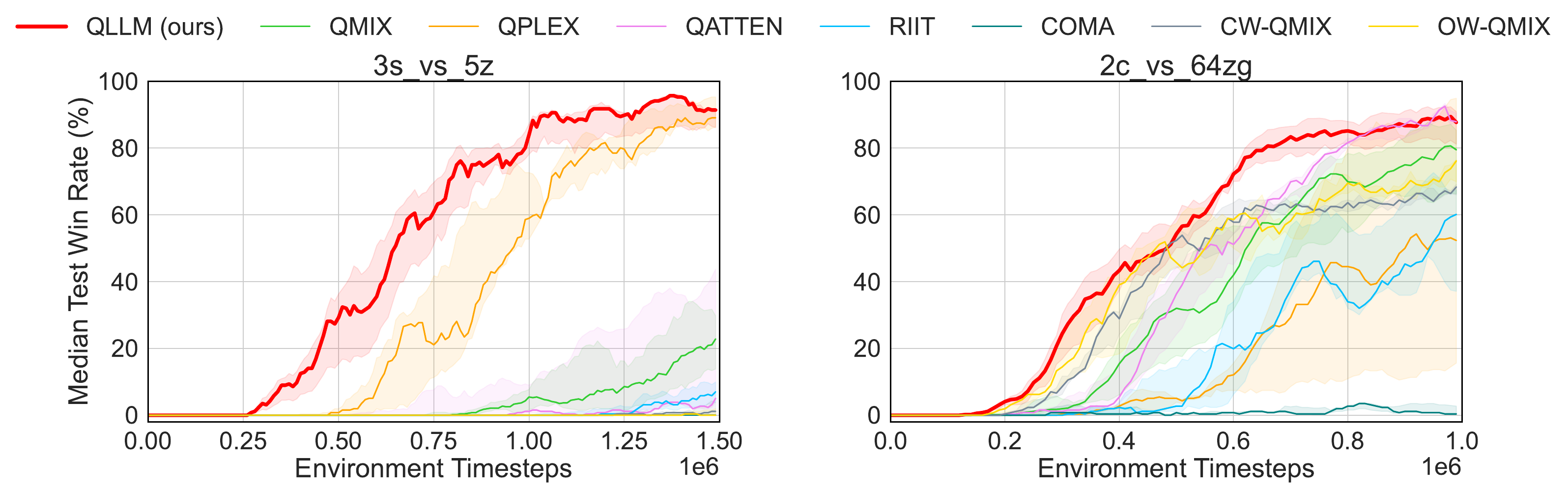}
    \caption{Performance comparison on SMAC maps.}
    \label{fig:smac}
\end{figure}

We evaluated QLLM on two hard SMAC maps: \textit{3s\_vs\_5z} and \textit{2c\_vs\_64zg}. As illustrated in Figure~\ref{fig:smac}, QLLM outperforms all baselines with significantly faster convergence. These results demonstrate that LLM-generated logic efficiently solves complex coordination tasks where traditional neural mixers typically struggle to optimize during the early stages of training.

\subsection{The Interpretability of TFCAF}

\begin{figure}[htbp]
    \centering
    \includegraphics[width=\textwidth]{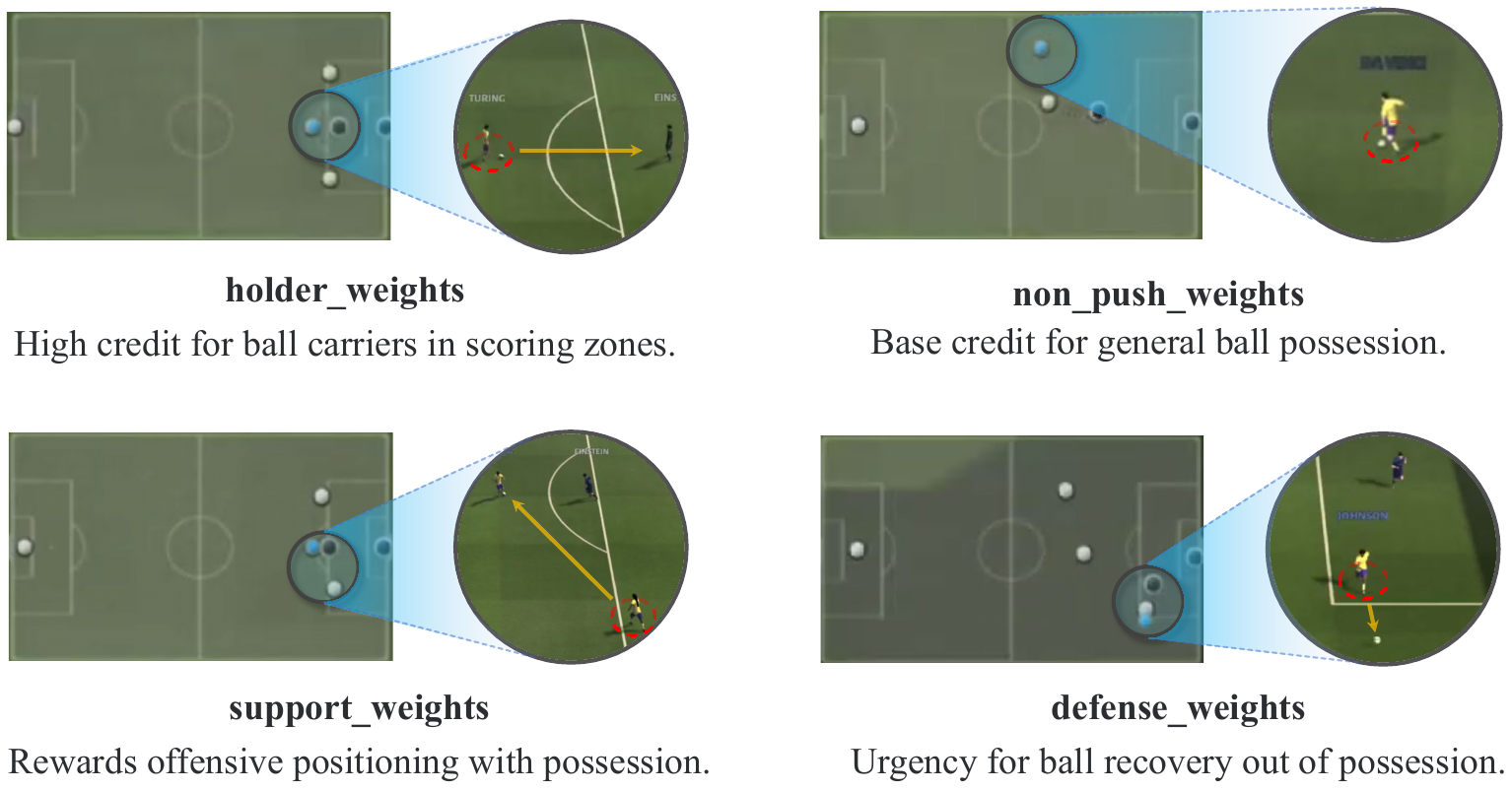}
    \caption{Visualization and Tactical Explanation of Weight Components in the TFCAF for \textit{academy\_3\_vs\_1\_with\_keeper} scenario.}
    \label{fig:TFCAF_GRF}
\end{figure}

To demonstrate the interpretability of QLLM, Listing~\ref{lst:grf_code} presents the complete generated TFCAF for the \textit{academy\_3\_vs\_1\_with\_keeper} scenario, while Figure~\ref{fig:TFCAF_GRF} provides a visualization and a detailed explanation of its weight components. The generated logic explicitly leverages physical factors, such as ball possession and goal proximity, to attribute credit. Unlike traditional neural mixing networks that operate as black boxes, TFCAF offers a transparent and human-readable mechanism, clearly revealing the underlying tactical rationale for credit assignment in complex environments.

\begin{listing}[htbp]
\centering
\caption{TFCAF for the \textit{academy\_3\_vs\_1\_with\_keeper} scenario in GRF.}
\label{lst:grf_code}
\begin{tcolorbox}[
    enhanced,
    colback=backcolour,
    colframe=gray!50!black,
    sharp corners=northeast,
    boxsep=2pt,
    top=1pt,
    bottom=2pt,
    left=2pt,
    right=2pt,
    before skip=5pt,
    after skip=5pt
]
\begin{lstlisting}[
    language=Python,
    basicstyle=\ttfamily\scriptsize, 
    numbers=none,
    aboveskip=0pt,
    belowskip=0pt,
    lineskip=1pt
]
def QLLMNetwork(agents_q, global_state):
    # Extract environment informations
    ball_x, ball_y = global_state[:, 88], global_state[:, 89]
    possession, holder_my_team=global_state[:,94:97],global_state[:,97:101]
    ball_goal_dist = torch.sqrt((ball_x-1.0)**2 + ball_y**2)
    my_team_has_ball = torch.argmax(possession, dim=1) == 0
    in_push_region = (ball_goal_dist>0.19) & (ball_goal_dist<0.99) & \                        my_team_has_ball
    agent_x, agent_y = global_state[:,[0,2,4,6]],global_state[:,[1,3,5,7]]
    agent_goal_dist = torch.sqrt((agent_x-1.0)**2 + agent_y**2)
    agent_ball_dist = torch.sqrt((agent_x-ball_x.unsqueeze(1))**2 + 
                                 (agent_y-ball_y.unsqueeze(1))**2)
    # Calculate weights components
    holder_weights = (1.0-ball_goal_dist).unsqueeze(1)*holder_my_team * \
                     in_push_region.unsqueeze(1)*10.0
    non_push_weights = holder_my_team*my_team_has_ball.unsqueeze(1) * \
                       (~in_push_region).unsqueeze(1)
    support_weights=(1.0/(1.0+agent_goal_dist))*(~holder_my_team.bool())* \
                     my_team_has_ball.unsqueeze(1)
    defense_weights = (1.0/(1.0+agent_ball_dist)) * \
                      (~my_team_has_ball).unsqueeze(1)
    weights = torch.softmax(holder_weights + non_push_weights + 
                            support_weights + defense_weights, dim=1)
    return (agents_q * weights).sum(dim=1, keepdim=True)
\end{lstlisting}
\end{tcolorbox}
\end{listing}

\section{Conclusion}
In this study, we propose QLLM, a method that uses LLMs to address the credit assignment problem in multi-agent reinforcement learning through coder-evaluator framework. This approach generates task-specific, training-free credit assignment functions, reducing hallucination and improving task understanding. Experiments on four standard MARL benchmarks show that QLLM consistently outperforms baselines and generalizes well across different mixing-network-based algorithms. Future work will explore broader applications and deployment in real-world multi-robot systems.



\bibliographystyle{splncs04}
\bibliography{ref} 

\newpage
\appendix
\section{LLM Prompts and Responses}
\label{sec:A}
\begin{tcolorbox}[title = {Coder Prompt}]
You are proficient in understanding tasks and writing Python code.
You should fully understand the multi-agent reinforcement learning tasks and environment, including the agents, action space, observation space, and global state space provided by the user.
Then, based on your understanding of the task and objectives, you need to write a function for multi-agent credit assignment, called QLLMNetwork, which is used to measure each agents' contribution to the collective reward.
The function input will be the individual Q-values of the agents and the global state, and the output will be the global Q-value.
Both input and output should be torch tensors.Individual Q-values represent each agent's expected contribution to the team's future rewards, while the global Q-value denotes the expected overall reward of the team as a whole.
The input Q-values will have a shape of torch.Size([batchsize, n\_agent]), the input global state will have a shape of torch.Size([batchsize, state\_dim]), and the output global Q-value should have a shape of torch.Size([batchsize, 1]). You must strictly follow the input and output dimension requirements!
In the credit assignment function you wrote, You need to extract the individual components of the local observation vectors of the individual agents.
You must accurately understand the meaning of each component in the global state space, and based on that, determine the weights for each Q-value.
You must understand the information and suggestions provided by the user in real-time to optimize the function you design, but strictly adhere to the function’s input and output requirements!\\
Note:\\
1.It must be a function, not a class. The parameters in the function are fixed and do not require training to calculate an accurate global Q-value. Do not use trainable layers such as nn.Linear.\\
2.You must accurately understand the meaning of each component in the global state space, without making any assumptions or guesses!\\
3.When calculating the weights, do not divide by values that may be close to zero. Be careful to check intermediate results, weights for invalid numbers!This is not solved just by adding a small number to the dividend.\\
4.The weights are usually a combination of several components, and when calculating the individual components of the weights, it is important to pay attention to the ratio between the components and the range of the value.
At the same time, the weights can be output through a softmax layer if it is necessary.\\
5.All tensor data should be computed on the GPU. Please write accurate code and strictly follow the template output below, without including any other explanatory text or code run sample.\\

def QLLMNetwork(agents\_q: torch.Tensor, global\_state: torch.Tensor) -$>$ torch.Tensor:\newline
\hbox{\ \ \ \ } \# You design this place\newline
\hbox{\ \ \ \ } return global\_q\\
\end{tcolorbox}

\begin{tcolorbox}[title = {Evaluator Prompt}]
You are an excellent evaluator of multi-agent reinforcement learning performance. I will attempt different functions in multi-agent reinforcement learning to solve the credit assignment problem.
This function is required to compute the global Q-values accurately without training. It must not use trainable layers such as nn.Linear.
Your task is to propose modifications to the function in brief plain text suggestions or return function selection results in an array format. Your suggestions must be in line with the task description and requirements, and should not be based on random assumptions or unrealistic scenarios. Please listen carefully to my subsequent instructions and do not provide additional answers!
When you output plain text suggestions, please ensure that you do not output any selections. When you output selections in an array format, please ensure that you do not output plain text suggestions.
Whether you output plain text suggestions or selections, consider if the function meets all the details and considerations specified in the task description. Do not use any information that has not been provided to you to answer!

\end{tcolorbox}

\begin{tcolorbox}[breakable, title = {Task Prompt of LBF}]
\newcommand{\taskdiff}[1]{\colorbox{gray!20}{\parbox{0.95\linewidth}{#1}}}

The Level-Based Foraging environment focus on the coordination of involved agents. The task for each agent is to navigate the grid-world map and collect items. Each agent and food item is assigned a level and food items are randomly scattered in the environment. In order to collect a food item, agents have to choose a pick-up action next to the food item. 

\textbf{1. Composition of the reward:} \\
\taskdiff{
\textbf{\textit{8x8-2p-2f:}} Such collection is only successful if 2 agents participate in the pickup at the same time. The task is an 8x8 grid-world with 2 agents and 2 food items. \\
\textbf{\textit{10x10-3p-3f:}} Such collection is successful if the sum of involved agents' levels is greater than or equal to the item level. The task is a 10x10 grid-world with 3 agents and 3 food items. \\
\textbf{\textit{15x15-4p-3f:}} Such collection is successful if the sum of involved agents' levels is greater than or equal to the item level. The task is a 15x15 grid-world with 4 agents and 3 food items.
} \\
At the same time the agents involved in the pickup must be one square above, below, left and right of the food item. Agents receive team reward equal to the level of the collected food item. Agents only receive non-zero team reward for picking up food items. \\

\textbf{2. States:} \\
Global state vector consists of 3 components: agents component, food items component and last step action. \\
\textbf{Total dimension:} \\
\taskdiff{
\textbf{\textit{8x8-2p-2f:}} 14 dimensions ($2\times3 + 2\times3 + 2$). \\
\textbf{\textit{10x10-3p-3f:}} 21 dimensions ($3\times3 + 3\times3 + 3$). \\
\textbf{\textit{15x15-4p-3f:}} 25 dimensions ($4\times3 + 3\times3 + 4$).
} \\
\textbf{Components detail:} \\

\textbf{(1) Agents Component} \\
\taskdiff{
\textbf{\textit{8x8-2p-2f:}} 2 agents $\times$ 3 features (agent\_i's x coordinate, y coordinate, and level). \\
\textbf{\textit{10x10-3p-3f:}} 3 agents $\times$ 3 features (agent\_i's x coordinate, y coordinate, and level). \\
\textbf{\textit{15x15-4p-3f:}} 4 agents $\times$ 3 features (agent\_i's x coordinate, y coordinate, and level).
} \\

\textbf{(2) Food Items Component} \\
\taskdiff{
\textbf{\textit{8x8-2p-2f:}} 2 food items $\times$ 3 features (food\_item\_i's x coordinate, y coordinate, and level). \\
\textbf{\textit{10x10-3p-3f:}} 3 food items $\times$ 3 features (food\_item\_i's x coordinate, y coordinate, and level). \\
\textbf{\textit{15x15-4p-3f:}} 3 food items $\times$ 3 features (food\_item\_i's x coordinate, y coordinate, and level).
} \\
When a food item has been picked up, then the respective values are replaced with (-1, -1, 0). \\

\textbf{(3) Last Actions Executed} \\
\taskdiff{
\textbf{\textit{8x8-2p-2f:}} 2 dims (This information is set to [-1, -1] when the environment is in the first step). \\
\textbf{\textit{10x10-3p-3f:}} 3 dims (This information is set to [-1, -1, -1] when the environment is in the first step). \\
\textbf{\textit{15x15-4p-3f:}} 4 dims (This information is set to [-1, -1, -1, -1] when the environment is in the first step).
} \\

\textbf{3. Actions:} \\
Each agent has six possible discrete actions: \\
\texttt{\{0: Noop, 1: Move North, 2: Move South, 3: Move West, 4: Move East, 5: Pickup\}} \\
While the first action corresponds to the action simply staying within its grid and the last action being used to pickup nearby food, the remaining four actions encode discrete 2D navigation. Picking up is only valid if the sum of the levels of the agents doing the action is greater than or equal to the level of the food. agents must be positioned one cell adjacent to the food.
\end{tcolorbox}

\begin{tcolorbox}[breakable, title = {Task Prompt of GRF}]
\newcommand{\taskdiff}[1]{\colorbox{gray!20}{\parbox{0.95\linewidth}{#1}}}

These tasks are offensive soccer games where the objective is to score goals by cooperating through passing and positioning to bring the ball as close as possible to the opponent's goal, located at coordinates (1.0, 0.0).

\textbf{1. Composition of the reward:} \\
The agents receive rewards consisting of two components: 
(1) Progress reward: Provides a team reward when an agent carries the ball closer to the opponent's goal (granted if ball possession is maintained and ball-to-goal distance is between 0.19 and 0.99). 
(2) Goal reward: A team reward given when a goal is successfully scored, with an additional bonus based on the shot distance (farther distance yields a higher bonus).

\taskdiff{
\textbf{\textit{3\_vs\_1:}} Our team controls 1 goalkeeper and 3 offensive players (4 agents). The opposing team has 1 goalkeeper and 1 defender. \\
\textbf{\textit{Counterattack:}} This is an 11-vs-11 game. We control 4 players (4 agents), while the rest of our team and all opponents are controlled by built-in AI. \\
\textbf{\textit{Pass\_and\_Shoot:}} Our team consists of 1 goalkeeper and 2 offensive players (3 agents). The opposing team has 1 goalkeeper and 1 defender.
} \\

\textbf{2. States:} \\
Each reinforcement learning agent receives a local observation of 115 dimensions. The global state vector is formed by concatenating the local observations of the controlled agents. \\

\textbf{Components detail:} \\

\textbf{(1) Players' Positions and Directions} \\
\taskdiff{
\textbf{\textit{3\_vs\_1:}} 0–7: Positions of our 4 players; 8–15: Directions of our 4 players; 16–19: Positions of 2 opponent players; 20–23: Directions of 2 opponent players. \\
\textbf{\textit{Counterattack:}} 0–21: Positions of all our 11 players; 22–43: Directions of all our 11 players; 44–65: Positions of all 11 opponent players; 66–87: Directions of all 11 opponent players. \\
\textbf{\textit{Pass\_and\_Shoot:}} 0–5: Positions of our 3 players; 6–11: Directions of our 3 players; 12–15: Positions of 2 opponent players; 16–19: Directions of 2 opponent players.
} \\

\textbf{(2) Ball and Possession Information} \\
88–90: Ball position (x, y, z); 91–93: Ball direction (x, y, z); 94–96: Ball possession (our team: [0, 1, 0], opponent team: [0, 0, 1], no one: [1, 0, 0]). \\

\textbf{(3) Agent Identification and Auxiliary Information} \\
\taskdiff{
\textbf{\textit{3\_vs\_1:}} 97–114: Invalid information. \\
\textbf{\textit{Counterattack:}} 97–107: 11-dimensional one-hot encoding indicating the agent's ID (used to retrieve corresponding player's data). 108–114: Invalid information. \\
\textbf{\textit{Pass\_and\_Shoot:}} 97–114: Invalid information.
} \\

\textbf{Total Global State Dimension:} \\
The global state vector is constructed by concatenating the local observations of the three controlled agents, resulting in a total of \\
\taskdiff{
\textbf{\textit{3\_vs\_1:}} 460 dimensions ($115 \times 4$). \\
\textbf{\textit{Counterattack:}} 460 dimensions ($115 \times 4$). \\
\textbf{\textit{Pass\_and\_Shoot:}} 345 dimensions ($115 \times 3$).
}
\end{tcolorbox}

\begin{tcolorbox}[breakable, title = {Task Prompt of MPE}]
\newcommand{\taskdiff}[1]{\colorbox{gray!20}{\parbox{0.95\linewidth}{#1}}}

\textbf{1. Composition of the reward:} \\
\taskdiff{
\textbf{\textit{Spread:}} We are playing Cooperative Navigation. In this task, $N$ agents are asked to reach $N$ landmarks (where $N \in \{6, 15, 25\}$). Each agent should move to get close to one landmark and avoid collision (distance $\le$ 0.3) with other agents. The agent-team are rewarded based on how far the closest agent is to each landmark (the reward equals the negative value of the sum of the minimum distances). At the same time, the agent-team are penalized if they collide each other (-1 for each collision). \\
\textbf{\textit{Adversary:}} In this environment, there is 1 adversary, 2 good agents, 2 landmarks. All agents observe the position of 2 landmarks and other agents. One of these 2 landmarks is the 'target landmark'. Good agents are rewarded based on how close the closest one of them is to the target landmark, but negatively rewarded based on how close the adversary is to the target landmark. The adversary want to get closer to the target landmark based on the agents' behavior, but it doesn't know which landmark is the target landmark. \\
\textbf{\textit{Tag:}} We are playing Predator Prey. In this task, 3 agents should cooperate to catch 1 prey. Agents move to get closer to the prey to catch it (distance $<$ 0.125). Prey is faster and tries to get out of the agents' clutches. Agents are slower and are rewarded for hitting preys (+10 for each collision). Obstacles will block the way.
} \\
At each step, each agent receives an observation array. \\

\textbf{2. State space:} \\
This observation is represented by an array with: \\
\taskdiff{
\textbf{\textit{Spread:}} 36 dimensions (for $N=6$), 90 dimensions (for $N=15$), or 150 dimensions (for $N=25$). \\
\textbf{\textit{Adversary:}} 8 dimensions. \\
\textbf{\textit{Tag:}} 16 dimensions.
} \\
The observation components are concatenated as: \\
\taskdiff{
\textbf{\textit{Spread:}} \texttt{concat([agent's velocity, agent's location, $N$ landmarks' relative locations, $N-1$ other agents' relative locations, communication code])}. \\
\textbf{\textit{Adversary:}} \texttt{concat([target landmark's relative locations, 2 landmarks' relative locations, other agents' relative locations (including adversary)])}. \\
\textbf{\textit{Tag:}} \texttt{concat([agent's velocity, agent's location, 2 obstacles' relative locations, 2 other agents' relative locations, 1 prey's relative locations, 1 prey's velocity])}.
} \\
The global state space is concatenated together from the observation space of all agents, resulting in: \\
\taskdiff{
\textbf{\textit{Spread:}} $36\times6$ (for $N=6$), $90\times15$ (for $N=15$), or $150\times25$ (for $N=25$) dimensions. \\
\textbf{\textit{Adversary:}} $8\times2$ dimensions. \\
\textbf{\textit{Tag:}} $16\times3$ dimensions.
} \\

\textbf{3. Action space:} \\
\texttt{[no\_action, move\_left, move\_right, move\_down, move\_up]}
\end{tcolorbox}

\begin{tcolorbox}[breakable, title = {Task Prompt of SMAC}]
\newcommand{\taskdiff}[1]{\colorbox{gray!20}{\parbox{0.95\linewidth}{#1}}}

\textbf{SECTION 1: CONTEXT AND OBJECTIVE} \\
You are an expert in Multi-Agent Reinforcement Learning (MARL), specifically focusing on the QLLM framework. Your task is to generate a PyTorch function named \texttt{QLLMNetwork(agents\_q, state)}. This function must act as a non-linear mixing network that takes individual agent Q-values with shapes: \\
\taskdiff{
\textbf{\textit{3s\_vs\_5z:}} agents\_q, shape [Batch, 3] \\
\textbf{\textit{2c\_vs\_64zg:}} agents\_q, shape [Batch, 2]
} \\
and the global state with shapes: \\
\taskdiff{
\textbf{\textit{3s\_vs\_5z:}} state, shape [Batch, 68] \\
\textbf{\textit{2c\_vs\_64zg:}} state, shape [Batch, 342]
} \\
to compute a scalar Global Q-value (shape [Batch, 1]). You must utilize the detailed features in the state vector to dynamically compute weights and biases for summing the agent Q-values. 

The agents receive rewards consisting of damage rewards (+1 per damage), kill rewards (+10 per kill), and a win bonus (+200). The return is scaled such that the maximum possible scaled return is 20. \\

\textbf{SECTION 2: ENVIRONMENT SPECIFICATION} \\
\taskdiff{
\textbf{\textit{3s\_vs\_5z:}} 3 Stalkers vs 5 Zealots. Stalkers (Allies) have 80 HP, 80 Shield, and 0.2142 attack range. Zealots (Enemies) have 100 HP, 50 Shield, and melee range (0.05). \\
\textbf{\textit{2c\_vs\_64zg:}} 2 Colossi vs 64 Zerglings. Colossi (Allies) have 200 HP, 150 Shield, and 0.25 attack range with splash damage. Zerglings (Enemies) have 35 HP, 0 Shield, and melee range (0.01).
} \\
Coordinates (X, Y) are normalized relative to the map center (range [-0.5, 0.5]) using a coordinate formula: (absolute\_coord - map\_center) / 28. \\

\textbf{SECTION 2.5: HEURISTIC REASONING GUIDE} \\
\taskdiff{
\textbf{\textit{3s\_vs\_5z:}} Stalkers are ranged units facing melee Zealots. Success depends on ``kiting"—maintaining a distance between 0.05 and 0.2142. Agents successfully kiting while attacking should receive higher weights. \\
\textbf{\textit{2c\_vs\_64zg:}} Colossi deal powerful splash damage. Success depends on maintaining distance from the massive swarm of 64 Zerglings. Credit should be weighted toward agents effectively engaging clusters of enemies while kiting. An agent being swarmed (multiple enemies at melee distance) should have its survival actions prioritized.
} \\
Common Rules: (1) Mask Dead Agent: If health is 0.0, weight must be forced to 0.0. (2) Efficiency: Agents with high weapon cooldown cannot attack; temporarily lower their weights unless they are drawing fire. (3) Shield Awareness: Protoss survivability depends on ``Health + Shield". Priority should be given to units with low total survivability. \\

\textbf{SECTION 3: DATA STRUCTURE} \\
\textbf{PART A: ALLY INFORMATION} \\
\taskdiff{
\textbf{\textit{3s\_vs\_5z:}} Indices 0-14. 3 agents, 5 features each: [Health, Cooldown, Rel\_X, Rel\_Y, Shield]. \\
\textbf{\textit{2c\_vs\_64zg:}} Indices 0-9. 2 agents, 5 features each: [Health, Cooldown, Rel\_X, Rel\_Y, Shield].
} \\

\textbf{PART B: ENEMY INFORMATION} \\
\taskdiff{
\textbf{\textit{3s\_vs\_5z:}} Indices 15-34. 5 enemies, 4 features each: [Health, Rel\_X, Rel\_Y, Shield]. \\
\textbf{\textit{2c\_vs\_64zg:}} Indices 10-201. 64 enemies, 3 features each: [Health, Rel\_X, Rel\_Y].
} \\

\textbf{PART C: LAST ACTIONS} \\
\taskdiff{
\textbf{\textit{3s\_vs\_5z:}} Indices 35-67. 3 agents, 11 actions each. Actions: 0:No-op, 1:Stop, 2-5:Move, 6-10:Attack Enemy 1-5. \\
\textbf{\textit{2c\_vs\_64zg:}} Indices 202-341. 2 agents, 70 actions each. Actions: 0:No-op, 1:Stop, 2-5:Move, 6-69:Attack Enemy 1-64.
} \\

Please write the QLLMNetwork function in PyTorch, ensuring all tensors are on state.device.
\end{tcolorbox}

\begin{tcolorbox}[breakable, title = {TFCAF template}] 

import torch\\
import torch.nn as nn\\
import torch.nn.functional as F\\
import numpy as np\\
class QLLMMixer(nn.Module):\\
\hbox{\ \ \ \ }def \_\_init\_\_(self):\\
\hbox{\ \ \ \ }\hbox{\ \ \ \ }super(QLLMMixer, self).\_\_init\_\_()\\
\hbox{\ \ \ \ }def forward(self, agents\_q, global\_state):\\
\hbox{\ \ \ \ }\hbox{\ \ \ \ }agents\_q = torch.squeeze(agents\_q)\\
\hbox{\ \ \ \ }\hbox{\ \ \ \ }a = agents\_q.shape[0]\\
\hbox{\ \ \ \ }\hbox{\ \ \ \ }b = agents\_q.shape[1]\\
\hbox{\ \ \ \ }\hbox{\ \ \ \ }agents\_q = agents\_q.reshape(-1, agents\_q.shape[-1])\\
\hbox{\ \ \ \ }\hbox{\ \ \ \ }global\_state = global\_state.reshape(-1, global\_state.shape[-1])\\
\hbox{\ \ \ \ }\hbox{\ \ \ \ }\#This is generated by LLM\\
\hbox{\ \ \ \ }\hbox{\ \ \ \ }return (global\_q\*agents\_q.shape[-1]).reshape(a, b, 1).cuda()\\

\end{tcolorbox}

\section{Additional Experiments}
\label{sec:B}

\subsection{Ablation Study on the Evaluator LLM}

While most LLM-enhanced reinforcement learning methods rely on a single model, our coder-evaluator framework leverages the collaboration between $M_{\text{coder}}$ and $M_{\text{evaluator}}$ to ensure logical precision. To isolate the impact of the evaluator, we evaluated a variant termed QLLM-C, which utilizes only the coder LLM for single-shot generation (effectively $K=1, T=1$). 

\begin{figure*}[htbp]
    \centering
    \includegraphics[width=\textwidth]{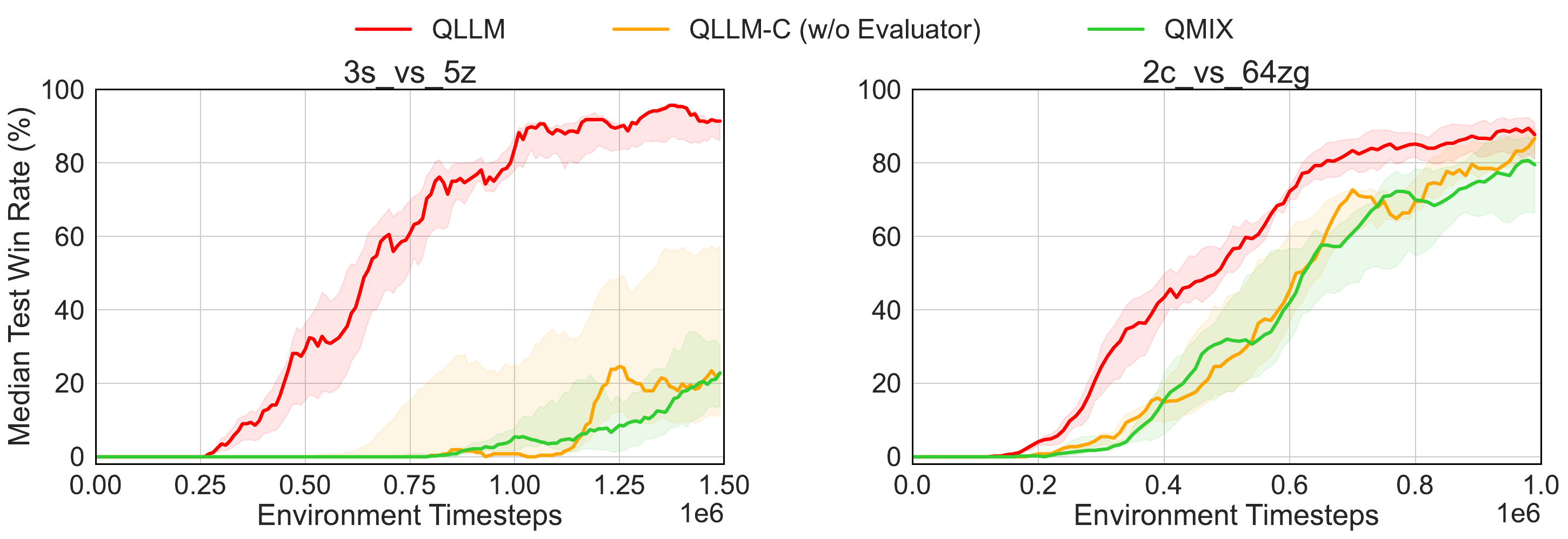}
    \caption{Performance comparison of QLLM and QLLM-C on SMAC maps. QLLM-C stands for the variant utilizing only the coder LLM for generation.}
    \label{fig:QLLM-C}
\end{figure*}

We conducted this ablation study on the SMAC \textit{3s\_vs\_5z} and \textit{2c\_vs\_64zg} scenarios. As illustrated in Figure~\ref{fig:QLLM-C}, QLLM-C exhibits a slight performance advantage over the QMIX baseline, suggesting that the zero-shot prior knowledge encoded in LLMs can capture basic coordination patterns. However, QLLM-C is significantly outperformed by the full QLLM framework in both scenarios. This performance gap demonstrates that $M_{\text{evaluator}}$ is essential for filtering out hallucinations and refining credit assignment logic through iterative feedback, especially in complex environments where high-fidelity reasoning is critical. These results validate the architectural necessity of our coder-evaluator framework.

\subsection{Ablation Study on Candidate Number and Iteration Rounds}

\begin{figure*}[htbp]
    \centering
    \includegraphics[width=\textwidth]{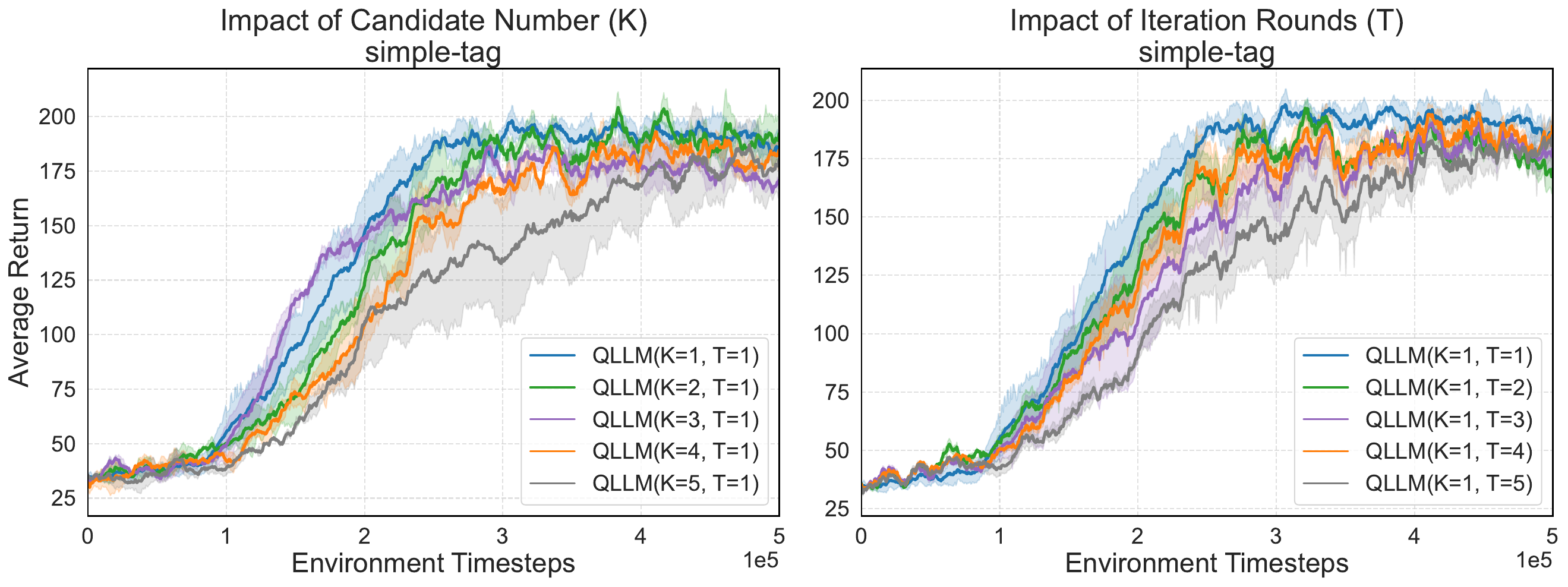}
    \caption{Ablation study of the number of candidates $K$ and iteration rounds $T$ on the \textit{pz-mpe-simple-tag} scenario.}
    \label{fig:KT}
\end{figure*}

To investigate the impact of the number of candidate TFCAFs per round ($K$) and the number of iteration rounds within the \textit{coder-evaluator} framework ($T$), we conducted parameter ablation studies on the MPE \textit{simple-tag} task. As illustrated in Figure~\ref{fig:KT}, the results reveal that in relatively simple scenarios like MPE, generating a single TFCAF without iterative optimization (i.e., $K=1, T=1$) is already sufficient to achieve near-optimal performance. 

However, this finding contrasts with our observations in complex environments such as SMAC. In fact, the QLLM-C variant (refer to Fig.~\ref{fig:QLLM-C}) effectively represents the $K=1, T=1$ configuration, and its performance in SMAC scenarios is significantly inferior to that of the standard $QLLM(K=3, T=3)$. We can conclude that for low-complexity tasks, minimal generation counts and iterations are adequate. In contrast, complex scenarios necessitate a larger candidate pool and multiple refinement cycles to synthesize TFCAFs with sophisticated and robust coordination logic.

\subsection{Compatibility Across Different LLMs}

\begin{figure*}[htbp]
    \centering
    \includegraphics[width=0.8\textwidth]{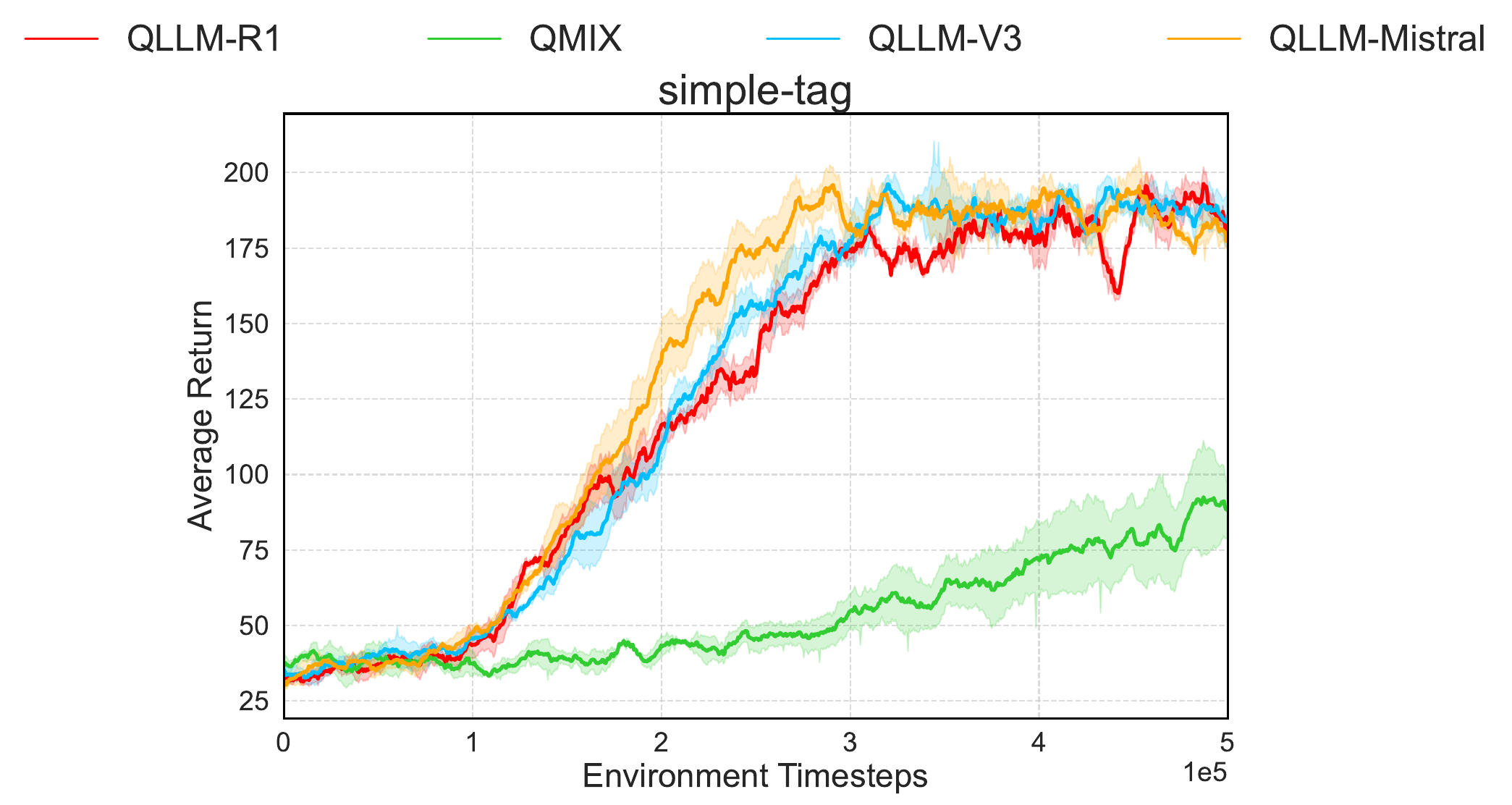}
    \caption{Performance comparison of QLLM variants using different mainstream LLMs on the MPE \textit{simple-tag} scenario.}
    \label{fig:diff}
\end{figure*}

To assess framework compatibility, we implemented QLLM using DeepSeek-R1~\cite{guo2025deepseek}, DeepSeek-V3\footnote{https://www.deepseek.com/}, and Mistral-Large\footnote{https://mistral.ai/} on the MPE \textit{simple-tag} task. As illustrated in Fig.~\ref{fig:diff}, QLLM consistently surpasses the QMIX baseline regardless of the underlying LLM architecture. Furthermore, the results show a positive correlation between LLM reasoning capabilities and RL performance; specifically, the Mistral-Large variant exhibits a slight performance margin over the others. These findings demonstrate that QLLM is a versatile framework capable of effectively leveraging the diverse reasoning strengths of various state-of-the-art LLMs.

\subsection{Comparison of Algorithms Training Costs}
\begin{figure*}[htbp]
		\centering
		\includegraphics[width=0.7\textwidth]{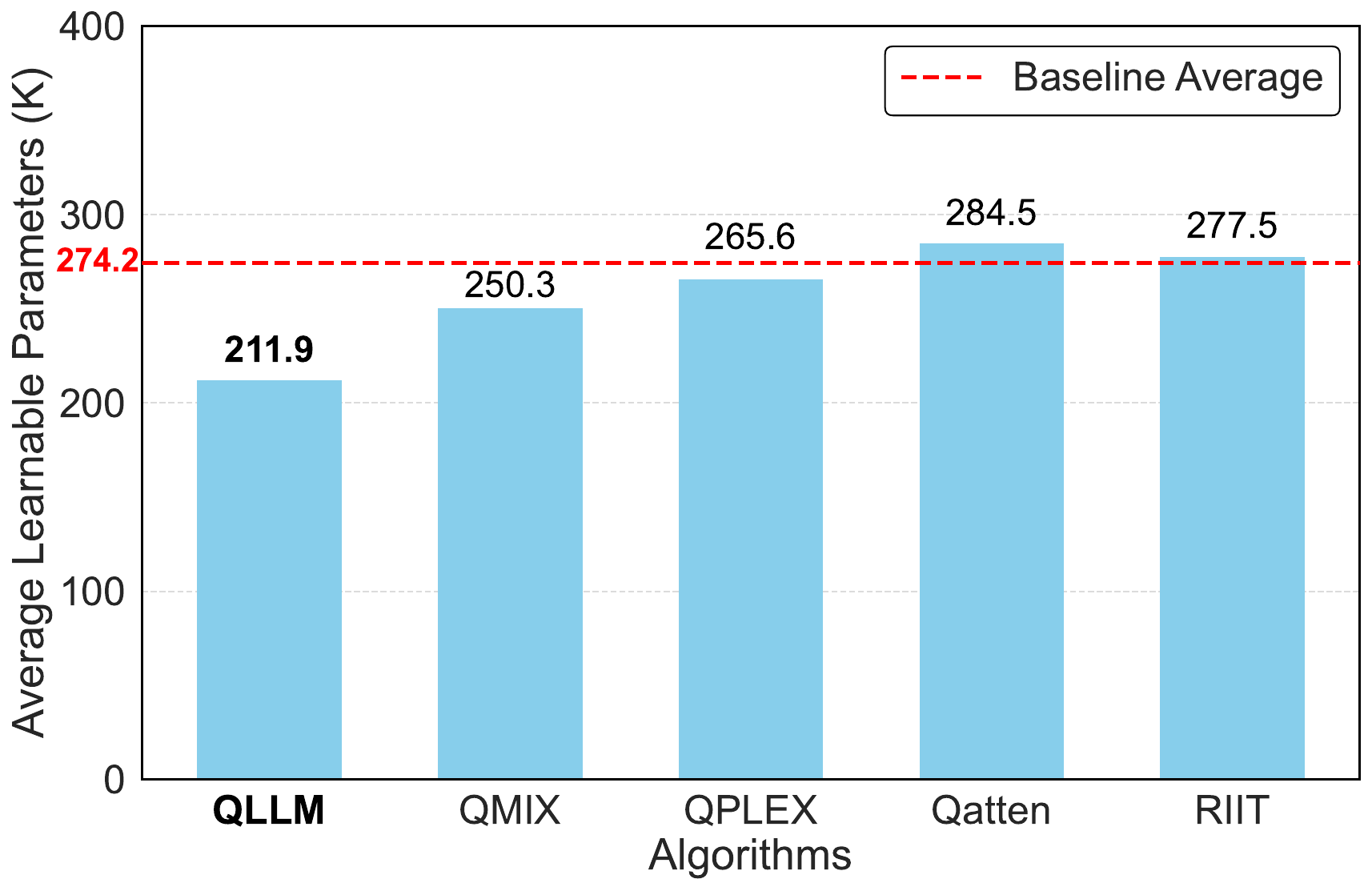}
		\caption{Comparison of the average number of learnable parameters across all tested environments.}
		\label{fig:parameters_num}
	\end{figure*}

As discussed in the methodology section, QLLM replaces the traditional neural mixing network with a deterministic functional expression, TFCAF. This architectural shift results in a significantly lower number of learnable parameters compared to standard value decomposition methods. Table~\ref{tab:cost} details the specific number of trainable parameters for each algorithm across the tested environments, while Figure~\ref{fig:parameters_num} provides a visual comparison of the average parameter counts. As illustrated, QLLM reduces the number of trainable parameters by approximately 13\% to 37\% relative to the baseline average. This substantial reduction simplifies the model structure and directly lowers the training cost, thereby enhancing the overall efficiency of the algorithm while maintaining superior performance.

Furthermore, we evaluated the operational efficiency of QLLM by comparing its average training duration against QMIX across these SMAC scenarios. Experimental results show that QLLM reduces the average training time per step from 161.00 ms to 104.18 ms (35.3\% reduction) and shortens the average total duration for 2 million training steps from approximately 9.05 hours to 5.38 hours (40.5\% reduction). Although QLLM includes an additional phase for TFCAF synthesis, one iteration of the code generation process (with \(n=3\)) takes only about 15 minutes, which is negligible compared to the several hours of training time saved. These findings demonstrate that QLLM not only achieves superior performance but also significantly optimizes training costs and time efficiency by reducing learnable parameters and accelerating the training procedure.

\begin{table}[h]
\renewcommand{\arraystretch}{1.3}
\centering
\caption{Number of learnable 0(in K) for each environment under different algorithms}
\label{tab:cost}
\resizebox{\linewidth}{!}{%
\begin{tabular}{lccccccc}
\toprule
\textbf{Environment / Algorithm} & \textbf{ QLLM } & \textbf{ QMIX } & \textbf{ QPLEX } & \textbf{ Qatten } & \textbf{ RIIT } & \textbf{\makecell{ Baseline\\Average }} & \textbf{\makecell{ Reduction \\ Rate }} \\
\midrule
\small\textit{pz-mpe-simple-spread}       & \textbf{105.22} & 140.42 & 154.76 & 191.11 & 191.18 & 166.74 & 36.88\% \\
\small\textit{pz-mpe-simple-tag}          & \textbf{102.28} & 112.81 & 128.01 & 129.03 & 132.20 & 125.91 & 18.78\% \\
\small\textit{pz-mpe-simple-adversary}    & \textbf{101.38} & 107.27 & 122.76 & 118.28 & 124.36 & 118.58 & 14.52\% \\
\small\textit{Foraging-2s-15x15-4p-3f}    & \textbf{103.17} & 111.82 & 127.37 & 121.96 & 130.54 & 123.28 & 16.30\% \\
\small\textit{Foraging-2s-8x8-2p-2f-coop} & \textbf{101.77} & 106.89 & 122.76 & 116.68 & 124.30 & 118.15 & 13.87\% \\
\small\textit{Foraging-2s-10x10-3p-3f}    & \textbf{102.66} & 109.74 & 125.45 & 120.04 & 127.91 & 121.19 & 15.28\% \\
\small\textit{academy\_3\_vs\_1\_with\_keeper} & \textbf{430.36} & 529.30 & 543.01 & 601.50 & 568.09 & 575.36 & 25.23\% \\
\small\textit{academy\_counterattack\_easy} & \textbf{430.36} & 529.30 & 543.01 & 601.50 & 568.09 & 575.36 & 25.23\% \\
\small\textit{academy\_pass\_and\_shoot\_with\_keeper} & \textbf{430.10} & 504.88 & 523.05 & 560.76 & 530.87 & 543.13 & 20.83\% \\
\bottomrule
\end{tabular}%
}
\end{table}

\subsection{Experimental Setup}
The experiments were conducted on computing nodes equipped with an Intel(R) Xeon(R) Platinum 8488C CPU and NVIDIA H100 PCIe GPUs with 80GB of VRAM. Our framework was implemented using Python 3.8.20, with PyTorch 2.1.0 serving as the deep learning backend. To ensure high-performance GPU acceleration, the system was configured with NVIDIA Driver version 570.195.03 and CUDA 12.8.

\section{Details of Environments and Baselines}
\label{sec:C}
\subsection{Baseline Algorithms}

\noindent \textbf{QMIX} (Rashid et al. 2020b) is a representative value decomposition method designed for cooperative MARL.  
It factorizes the joint action-value function into individual agent utilities through a mixing network that enforces a monotonic constraint, ensuring that maximizing local utilities also maximizes the global return.  
The mixing network parameters are generated by a hypernetwork conditioned on the global state, enabling state-dependent flexible mixing.

\noindent \textbf{QPLEX} (Wang et al. 2020) builds on QMIX by introducing a duplex dueling architecture to overcome the limited expressiveness caused by strict monotonicity.  
It splits the value factorization into a main (advantage) stream and a complementary stream, allowing more flexible interactions while maintaining decentralized policies.  
This design expands the class of joint value functions that can be represented under the CTDE paradigm.

\noindent \textbf{Qatten} (Yang et al. 2020) enhances value decomposition by embedding a multi-head attention mechanism into the mixing network.  
This mechanism adaptively adjusts the importance of each agent’s local utility based on the global state context, enabling the model to focus on the most relevant agents or features when computing the global Q-value.

\noindent \textbf{RIIT} (Hu et al. 2021) tackles the multi-agent credit assignment problem by modeling implicit influence relationships among agents.  
It proposes a reward interpolation technique that redistributes the global team reward across agents, considering both immediate and long-term interdependencies, which helps reduce misattribution and improves coordination performance.

\noindent \textbf{COMA} (Foerster et al. 2018) is a centralized actor-critic method designed for cooperative MARL, with a focus on addressing the multi-agent credit assignment problem.
It employs a centralized critic that estimates a counterfactual baseline for each agent by marginalizing out its action while keeping others fixed, enabling the computation of an advantage function that isolates each agent’s contribution.
This counterfactual reasoning helps reduce variance in the policy gradient and promotes coordinated behaviors.

\noindent \textbf{Weighted QMIX} (Rashid et al. 2020a) further generalizes QMIX by learning an additional weighting factor for each agent’s contribution to the global Q-value.  
By relaxing the monotonicity assumption through these learnable weights, it improves the accuracy of credit assignment in tasks with more complex cooperative dependencies.

\noindent \textbf{MASER} (Jeon et al. 2022) is a value-decomposition-based MARL algorithm designed to improve credit assignment under sparse and delayed rewards.
It introduces subgoals automatically generated from the experience replay buffer, which are used to construct intrinsic rewards that guide agents toward meaningful intermediate states. By augmenting standard Q-learning with subgoal-conditioned value estimation, MASER enhances exploration efficiency and coordination while remaining compatible with centralized training and decentralized execution.

\subsection{Code Repository}
The baseline algorithms in this experiment are implemented based on the \textit{epymarl}\footnote{https://github.com/uoe-agents/epymarl}, \textit{pymarl}\footnote{https://github.com/oxwhirl/pymarl}, and \textit{pymarl2}\footnote{https://github.com/hijkzzz/pymarl2} codebases. Specifically, we evaluated the MPE and LBF environments using the \textit{epymarl} library. The StarCraft Multi-Agent Challenge (SMAC) was tested utilizing the original \textit{pymarl} framework, while Google Research Football (GRF) was evaluated with \textit{pymarl2}. This setup ensures experimental rigor and facilitates fair and controlled comparisons across diverse cooperative tasks.

\subsection{Tasks Details}

\begin{figure}[htbp]
\centering
\includegraphics[width=0.26\textwidth]{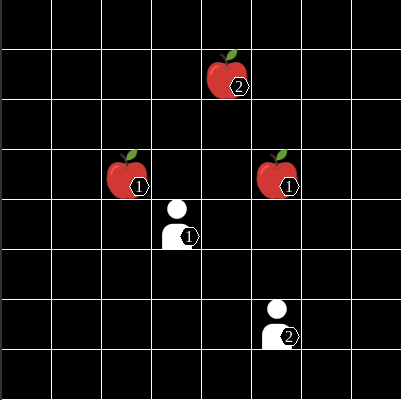}
\hspace{1pt}
\includegraphics[width=0.45\textwidth]{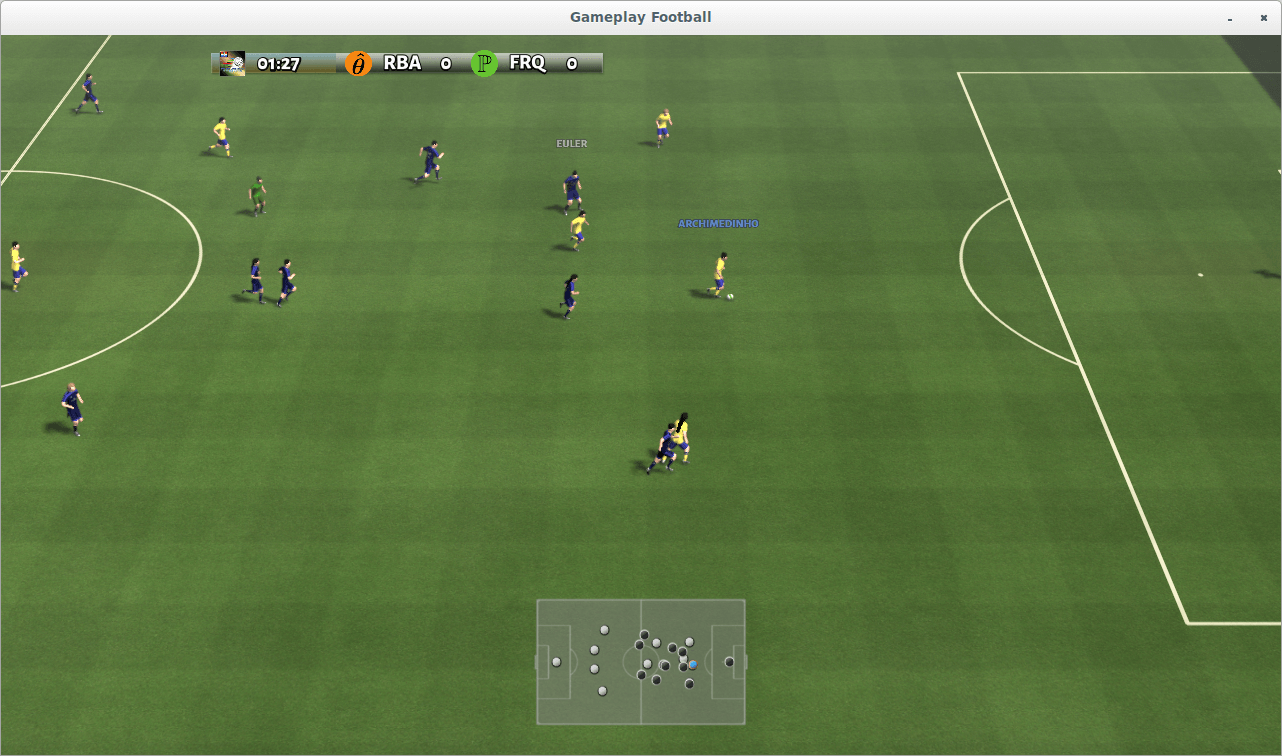}
\hspace{2pt} 
\includegraphics[width=0.26\textwidth]{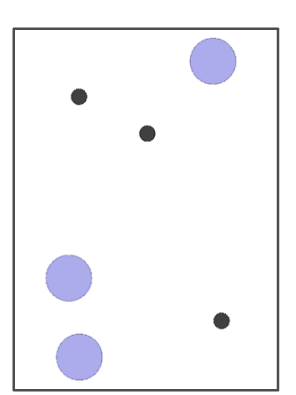}
\hspace{1pt} 
\includegraphics[width=0.45\textwidth]{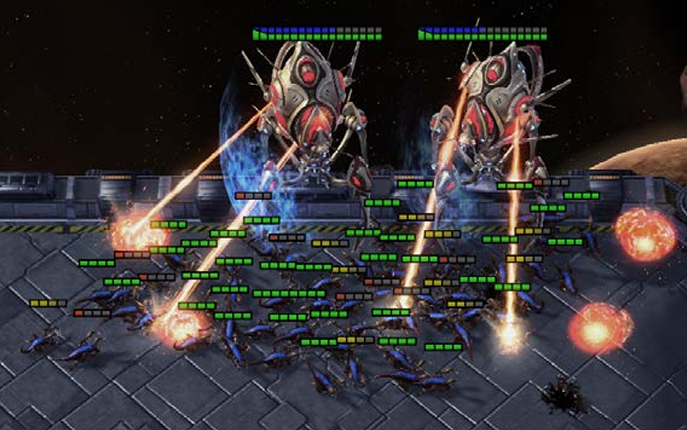}
\caption{Environment visualizations of Level-Based Foraging (LBF), Google Research Football (GRF), Multi-Agent Particle Environments (MPE) and StarCraft Multi-Agent Challenge (SMAC).}
\end{figure}

\subsubsection{Level-Based Foraging}
The Level-Based Foraging environment consists of tasks focusing on the coordination of involved agents. The task for each agent is to navigate the grid-world map and collect items. Each agent and item is assigned a level and items are randomly scattered in the environment. In order to collect an item, agents have to choose a certain action next to the item. However, such collection is only successful if the sum of involved agents’ levels is equal or greater than the item level. Agents receive reward equal to the level of the collected item. The tasks that were selected are denoted first by the grid-world size (e.g. 15 × 15 means a 15 by 15 grid-world). Then, the number of agents is shown (e.g. “4p” means four agents/players), and the number of items scattered in the grid (e.g. “3f” means three food items). We also have some special flags, the “2s” which denotes partial observability with a range of two squares. In those tasks, agents can only observe items or other agents as long as they are located in a square of size 5x5 centred around them. Finally, the flag “c” means a cooperative-only variant were all items can only be picked up if all agents in the level attempt to load it simultaneously. Since the global state space of the original environment is stitched together from the observation vectors of individual agents, it contains a lot of duplicated information and some environmental information is lost. Therefore, we redesigned a global state space which contains all the environment information while reducing the size of the dimension. Specifically, it includes the coordinates and levels of individual intelligences and things, as well as the actions executed by each agent last step.

\begin{table}[h]
\renewcommand{\arraystretch}{1.2}  
\centering
\caption{Details of LBF tasks}
\begin{tabular}{lcccc}
\toprule
\textbf{LBF Tasks} & \textbf{ N\_agents } & \textbf{ State Shape } & \textbf{ Action Num }\\
\midrule
\textit{Foraging-2s-15x15-4p-3f}      & 4 & 25 & 6\\
\textit{Foraging-2s-8x8-2p-2f-coop}      & 2 & 14 & 6\\
\textit{Foraging-2s-10x10-3p-3f}      & 3 & 21 & 6\\
\bottomrule
\end{tabular}
\end{table}

\subsubsection{Google Research Football}
The Google Research Football environment is a challenging reinforcement learning platform designed to simulate football matches in a highly realistic and strategic setting. Agents control individual players and must learn to cooperate, plan, and execute complex behaviors to achieve objectives such as scoring goals or defending against opponents. The environment provides various predefined scenarios as well as full 11-versus-11 matches, supporting both single-agent and multi-agent training. In the Google Research Football environment, the action space consists of a discrete set of predefined high-level actions such as moving, passing, shooting, and sprinting, allowing agents to focus on strategic decision-making rather than low-level motor control. The \textit{Simple115} observation space is a fixed-size vector with 115 features, providing essential information about the game state, including the positions and velocities of the ball and players, the directions of movement, and other key attributes necessary for learning effective policies.

\begin{table}[h]
\renewcommand{\arraystretch}{1.2} 
\centering
\caption{Details of GRF tasks}
\begin{tabular}{lcccc}
\toprule
\textbf{GRF Tasks} & \textbf{ N\_agents } & \textbf{ State Shape } & \textbf{ Action Num }\\
\midrule
\textit{academy\_3\_vs\_1\_with\_keeper}              & 4 & 460 & 19\\
\textit{academy\_counterattack\_easy}                 & 4 & 460 & 19\\
\textit{academy\_pass\_and\_shoot\_with\_keeper}      & 3 & 345 & 19\\
\bottomrule
\end{tabular}
\end{table}

\subsubsection{Multi-Agent Particle Environments}
The Multi-Agent Particle Environments are a widely-used set of communication-oriented environments where particle agents can move, communicate, observe each other, exert physical forces, and interact with fixed landmarks within a 2D continuous space[1]. These environments are highly configurable and serve as standard benchmarks for evaluating coordination, competition, and communication in MARL.\\
\textit{\textbf{Simple-Spread:}} In this task, three agents are trained to move to three landmarks while avoiding collisions with each other. All agents receive their velocity, position, relative position to all other agents and landmarks. The action space of each agent contains five discrete movement actions. Agents are rewarded with the sum of negative minimum distances from each landmark to any agent and a additional term is added to punish collisions among agents.\\  
\textit{\textbf{Simple-Adversary:}} In this task, two cooperating agents compete with a third adversary agent. There are two landmarks out of which one is randomly selected to be the goal landmark. Cooperative agents receive their relative position to the goal as well as relative position to all other agents and landmarks as observations. However, the adversary agent observes all relative positions without receiving information about the goal landmark. All agents have five discrete movement actions. Agents are rewarded with the negative minimum distance to the goal while the cooperative agents are additionally rewarded for the distance of the adversary agent to the goal landmark. Therefore, the cooperative agents have to move to both landmarks to avoid the adversary from identifying which landmark is the goal and reaching it as well. For this competitive scenario, we use a fully cooperative version where the adversary agent is controlled by a pretrained model.\\
\textit{\textbf{Simple-Tag:}} In this task, three cooperating predators hunt a forth agent controlling a faster prey. Two landmarks are placed in the environment as obstacles. All agents receive their own velocity and position as well as relative positions to all other landmarks and agents as observations. Predator agents also observe the velocity of the prey. All agents choose among five movement actions. The agent controlling the prey is punished for any collisions with predators as well as for leaving the observable environment area (to prevent it from simply running away without needing to learn to evade). Predator agents are collectively rewarded for collisions with the prey. We employ a fully cooperative version of this task with a pretrained prey agent.

\begin{table}[h]
\renewcommand{\arraystretch}{1.2}  
\centering
\caption{Details of MPE tasks}
\begin{tabular}{lcccc}
\toprule
\textbf{MPE Tasks} & \textbf{ N\_agents } & \textbf{ State Shape } & \textbf{ Action Num }\\
\midrule
\textit{pz-mpe-simple-spread}      & 6 & 216 & 5\\
\textit{pz-mpe-simple-tag}      & 3 & 48 & 5\\
\textit{pz-mpe-simple-adversary}      & 2 & 16 & 5\\
\bottomrule
\end{tabular}
\end{table}

\subsubsection{StarCraft Multi-Agent Challenge}
The StarCraft Multi-Agent Challenge is a high-dimensional micro-management benchmark based on the StarCraft II engine. In this environment, agents must learn to cooperate with each other and execute sophisticated combat strategies (such as focus fire and kiting) to defeat enemy units and achieve victory. To evaluate the performance of our algorithm in complex multi-agent cooperative scenarios, we selected two challenging maps categorized as hard difficulty: \textit{3s\_vs\_5z} and \textit{2c\_vs\_64zg}. These scenarios require sophisticated micromanagement and coordination to handle heterogeneous units or large-scale enemy swarms. The specific parameters of these two maps are detailed in Table~\ref{tab:smac}.

\begin{table}[h!]
\renewcommand{\arraystretch}{1.2}
\centering
\caption{Details of the SMAC scenarios.}
{
\begin{tabular}{lcccc}
\toprule
\textbf{SMAC Tasks } & \textbf{  Ally Units  } & \textbf{  Enemy Units  } & \textbf{  State Shape  } & \textbf{  Action Num  } \\
\midrule
\textit{3s\_vs\_5z} & 3 Stalkers & 5 Zealots & 68 & 11 \\
\textit{2c\_vs\_64zg} & 2 Colossi & 64 Zerglings & 342 & 70 \\
\bottomrule
\end{tabular}
}
\label{tab:smac}
\end{table}

\section{Proof of the QLLM Decomposition Formula}
\label{sec:appendix_proof}

In this section, we provide a formal derivation of the decomposition formulation used in QLLM (Equation~\eqref{eq:qllm}). We aim to prove that the joint action-value function $Q_{\text{tot}}(s, \boldsymbol{a})$ can be represented as a state-dependent linear combination of individual utility functions $Q_i(\tau^i, a^i)$ plus a state-dependent bias term $f_b(s)$. 

Assuming the joint value function $Q_{\text{tot}}$ is differentiable with respect to the individual $Q$-values, we can perform a multivariate Taylor expansion of $Q_{\text{tot}}$ at the state $s$ as a function of the utility vector $\boldsymbol{Q} =(Q_1, \dots, Q_n)$:
\begin{equation}
Q_{\text{tot}}(s, \boldsymbol{Q}) = \text{const}(s) + \sum_{i=1}^n \mu_i(s) Q_i + \sum_{i,j} \mu_{ij}(s) Q_i Q_j + \mathcal{O}(Q^3),
\label{eq:proof_taylor}
\end{equation}
where 
\[ 
\mu_i(s) = \frac{\partial Q_{\text{tot}}}{\partial Q_i} \quad \text{and} \quad \mu_{ij}(s) = \frac{1}{2} \frac{\partial^2 Q_{\text{tot}}}{\partial Q_i \partial Q_j}. 
\]
To analyze the local behavior near an optimal joint action $\boldsymbol{a}_0 = (a_0^1, \dots, a_0^n)$, we note that the gradient $\nabla_{\boldsymbol{a}} Q_{\text{tot}}$ vanishes at $\boldsymbol{a}_0$. Since 
\[ 
\frac{\partial Q_{\text{tot}}}{\partial a^i} = \frac{\partial Q_{\text{tot}}}{\partial Q_i} \frac{\partial Q_i}{\partial a^i} = 0, 
\]
and assuming that each agent has a non-trivial impact on the team ($\mu_i \neq 0$), the local gradient of each agent's utility must also vanish: 
\[ 
\frac{\partial Q_i}{\partial a^i}(a_0^i) = 0. 
\]
Consequently, each individual utility function can be approximated by its local quadratic expansion: 
\[ 
Q_i(a^i) \approx \alpha_i(s) + \beta_i(s)(a^i - a_0^i)^2. 
\]

The key to linearizing the formulation lies in the treatment of the second-order cross terms $Q_i Q_j$. Substituting the quadratic approximation into these terms yields:
\begin{align}
Q_i Q_j &\approx[\alpha_i + \beta_i(a^i - a_0^i)^2][\alpha_j + \beta_j(a^j - a_0^j)^2] \nonumber \\
&= \alpha_i \alpha_j + \alpha_i \beta_j(a^j - a_0^j)^2 + \alpha_j \beta_i(a^i - a_0^i)^2 + \textit{higher-order terms} \nonumber \\
&\approx \alpha_i \alpha_j + \alpha_i (Q_j - \alpha_j) + \alpha_j (Q_i - \alpha_i) \nonumber \\
&= \alpha_j Q_i + \alpha_i Q_j - \alpha_i \alpha_j.
\label{eq:proof_linearization}
\end{align}

Equation~\eqref{eq:proof_linearization} demonstrates that the product of two utilities near the optimum can be effectively represented as a linear combination of $Q_i$ and $Q_j$ plus a state-dependent constant. By substituting this result into the Taylor expansion in Equation~\eqref{eq:proof_taylor} and grouping all terms associated with each $Q_i$, we obtain:

\begin{equation}
\begin{split}
Q_{\text{tot}}(s, \boldsymbol{a}) &\approx \sum_{i=1}^n \underbrace{\left[ \mu_i(s) + 2 \sum_{j \neq i} \mu_{ij}(s) \alpha_j(s) + \dots \right]}_{f_{w}^i(s)} Q_i \\
&\quad + \underbrace{\left[ \text{const}(s) - \sum_{i,j} \mu_{ij}(s) \alpha_i \alpha_j + \dots \right]}_{f_{b}(s)}.
\end{split}
\end{equation}

This derivation confirms that the formulation $Q_{\text{tot}}(s, \boldsymbol{a}) = \sum f_w^i(s) Q_i + f_b(s)$ is a mathematically sound approximation for any complex value-based mixing mechanism that satisfies the Individual-Global-Max (IGM) principle. While traditional methods like QMIX or Qatten utilize hypernetworks to estimate these coefficients, QLLM leverages the logical reasoning of LLMs to generate the functions $f_w^i(s)$ and $f_b(s)$ directly, which enhances both interpretability and parameter efficiency.

\end{document}